\documentclass{aa}
\usepackage{graphicx}

\begin{document}

\title{Star formation in accretion discs : \\ from the Galactic center to active galactic nuclei}

\author{S. Collin\inst{1} \and J.-P. Zahn\inst{1}}

\offprints{Suzy Collin (suzy.collin@obspm.fr)}

\institute{$^1$ LUTH, Observatoire de Paris, CNRS, Universit\'e Paris Diderot ; 5 Place Jules Janssen, 92190 
Meudon, France 
}

\titlerunning{Star formation in accretion discs}
\authorrunning{S. Collin, J.-P. Zahn}

\abstract
{Keplerian accretion discs around massive black holes (MBHs) are gravitationally unstable beyond a few hundredths of a parsec, and they should collapse to form stars. It has indeed been shown recently that an accretion/star formation episode took place a few million years ago in the Galactic center (GC). This raises the question of how the disc can survive in AGN and quasars and continue to transport matter towards the black hole. }{We study the accretion/star formation process in quasars and AGN with one aim in mind: to show that a spectrum similar to the observed one can be produced by the disc.}{We compute models of stationary accretion discs that are either continuous or clumpy. Continuous discs must be maintained in a state of marginal stability so that the rate of star formation remains modest and the disc is not  immediately destroyed.
The disc then requires additional heating and additional transport of angular momentum. In clumpy discs, the momentum transport is provided by cloud interactions. }{Non-viscous heating can be provided by stellar illumination, but in the case of continuous discs, even momentum transport by supernovae is insufficient for sustaining a marginal state, except at the very periphery of the disc. In clumpy discs it is possible to account for the required accretion rate through interactions between clouds, but this model is unsatisfactory because its parameters are tightly constrained without any physical justification.  
}{Finally one must appeal to non-stationary discs with intermittent accretion episodes like those that occurred in the GC, but such a model is probably not applicable either to luminous high redshift quasars or to radio-loud quasars.} 


\maketitle

\section{``Historical" introduction}

It is well-known that Keplerian accretion discs around massive black holes (MBHs) are self-gravitating  at distances larger than about 1000 Schwarzschild radii (10$^3R_{\rm Schw}$). Consequently they become gravitationally unstable to fragmentation on a short timescale and can collapse to form stars (Paczynski 1978; Kolykhalov \& Sunyaev 1980; Lin \& Pringle 1987; Shlosman, Frank \& Begelman 1989; Shlosman \& 
Begelman 1989;  Hur\'e 1998; Collin \&
Zahn 1999a, 1999b; Gammie 2001; Goodman 2003; Thompson et al. 2005, and several subsequent papers discussing the Galactic center). Actually, Shlosman \& 
Begelman (1989) showed that unless  
they can be maintained in a hot or highly turbulent state, the discs 
should transform
rapidly into stellar systems  unable to fuel active galactic nuclei (AGN). Since AGN discs live for long times (10$^{7-8}$ years), it was mandatory to understand how they can avoid such collapse. 

This is why in a previous paper (Collin \& Zahn 1999b, hereafter referred as CZ99), we developed an exploratory model to show that in the presence of star formation, the accretion disc would be heated by the stars and could thus  be 
maintained 
in a marginally stable state where the star formation rate is strongly reduced and destruction of the gaseous disc is avoided. Besides, we predicted that the stellar mass function would be top-heavy due to continuing accretion onto the forming stars, giving rise to massive stars. 
We postulated that the momentum transport was caused by supernova explosions. We mentioned that this model could be relevant for explaining the presence of hot massive stars in the central parsec close to Sgr A*. Our main concern in that paper was to show that a rate of star formation {\it consistent} with the rate of supernovae explosions could sustain the accretion rate. In this sense it was restrictive. We found a solution, but only in a relatively narrow range of radius (about a decade), and relatively far from the MBH (at about 10$^6$ $R_{\rm Schw}$).

At that time there was no proof that stars, particularly massive stars, exist very close to SMBs. Only recently have starbursts been discovered in Seyfert nuclei on the 10 parsec scale in two Seyfert nuclei (Davies et al. 2007).  In 1999, a few groups of young stars were observed in the nuclear regions of some nearby AGN, but not closer than a few tens of parsecs. At this distance, the disc is  globally unstable as its mass becomes comparable to that of the MBH, and gravitational torques produced by bars and spiral waves mix the gas  copiously and can transport the mass. 

Concerning the Galactic center (GC), ideas have evolved considerably since the publication of CZ99, because many young massive stars have been discovered in the central parsec (Eckart, Ott \&  Genzel 1999). At first, their presence was considered very puzzling, because it was thought that the strong tidal field of the MBH should inhibit star formation. Gerhard (2001) thus proposed that these stars were born farther out in a star cluster and later drifted close to the MBH by dynamical friction with background stars. 
Several recent discoveries have made this scenario  implausible. Nayakshin \& Sunyaev (2005) showed that the maximum possible number of young low-mass stars (YSO), deduced from the comparison of Chandra observations of X-ray emission near Sgr A* with the Orion Nebula, is too small for the cluster inspiral model. In addition, they showed that the initial mass function (IMF) of the stars near the GC is top-heavy. Therefore
the ``in situ accretion disc scenario" gained credit and gave rise to various discussions (Levin \&
Beloborodov 2003; Goodman
2003; Milosavljevic \& Loeb 2004; Nayakshin 2004; Nayakshin
\& Cuadra 2005; Nayakshin 2006 and 2007; Levin 2007; Nayakshin et al. 2007). It seems clear now that star formation has taken place in the GC in an accretion disc and that the duration of this episode was rather short. 

However, one is still faced with a problem concerning AGN: no answer has been found so far on how star formation can be limited to a small amount or even be avoided during long periods of time in the accretion discs of AGN. As mentioned above, one possibility is that a stationary disc is maintained in a marginally stable state for a long time. A second possibility is that the disc is not stationary, and like the GC, is prone to transient accretion episodes.  We explore these two possibilities here.

Actually we tackle the problem in a ``reverse" way compared to what has been done so far. Because it was claimed (Sirko \& Goodman 2003) that stellar heating is unable to maintain an accretion disc in a marginally stable state without producing a spectrum very different from that of  a typical AGN, our basic assumption is that the disc emits a spectral energy distribution (SED) in agreement with the observations of AGN. We first assume that the disc is stationary and then check whether an acceptable amount of transport of angular momentum needs to be added  to the standard viscous one to maintain the disc in a marginally stable state. We conclude that supernovae do not cause enough angular momentum transport in a large fraction of the disc. We also briefly examine whether the clumpiness that would develop naturally in a gravitationally unstable disc could provide the required ``viscosity" through collisions of clumps, and we show that it could indeed be the case, but only if the size of the clumps is restricted to a narrow range of values.  We conclude that one probably has to seek a solution in the framework of non stationary discs, and we look whether they can be accommodated with the observations of AGN.

In the next section we discuss the lifetime of the episodes of star formation {\it and} accretion both in the GC and in luminous AGN. In Section 3 we recall the theoretical basis of stationary accretion discs and of the star formation mechanism. Results concerning typical quasars and Seyfert nuclei are  given in Section 4. In Section 5 we develop a simple model for accretion in a stationary clumpy disc, and finally, the last section is devoted to some considerations about non stationary accretion discs.

\section{How long should the accretion / star formation episodes last?}

\subsection{The case of the Galactic center and of low luminosity AGN}

   For the GC, our present state of knowledge has been summarized in Morris \& Nayakshin (2007). For the stars, it is now known that most of the young massive stars in the central parsec of the GC are located in a relatively thin disk (aspect ratio $\sim$ 0.1) rotating clockwise, while a few others are found in a diffuse disk rotating counter clockwise and inclined with respect to the first one (Levin \& Beloborodov 2003; Genzel et al. 2003; Paumard et al. 2006).  From the spectral study of these stars, Paumard et al. (2006) have deduced the composition of the clockwise and counter clockwise discs. Using the evolutionary tracks developed in Geneva with different IMF, they found that the IMF is flatter than the Salpeter mass function, confirming a formation in a non-common medium and that the proportions of different stellar types in the two discs can be well fitted by an episode of star formation taking place about 5 million years ago. They excluded a continuous formation process lasting more than two million years on the basis of the large number of WR stars.  This implies that  the gaseous disc around Sgr A* in which star formation took place has disappeared or has been unable to form new stars for 5 million years. 

Concerning the gaseous component, the inner 3 parsecs around Sgr A*  constitute a highly complex
region. There are infalling streamers of atomic and ionized gas (the ``minispiral") in the inner parsec, embedded into a ring of molecular gas (the ``circum-nuclear disc", CND) peaking at 1.5 pc, made up of about 30 dense clouds with masses on the order of 10$^4$ M$_{\odot}$ each and core densities of 10$^{7-8}$ particules per cc (Christopher et al. 2005).  The CND rotates at 110 km s$^{-1}$
 with some deviations, and its structure seems consistent with a warped
disc in Keplerian rotation around the MBH. The minispiral seems to be the inner edge of the CND, and both structures are in strong interaction. 

Thus the inner regions located at a distance smaller than one parsec must permanently receive gas from the molecular clouds. The inflow rate deduced from the atomic gas has been estimated as 3 10$^{-2}$ M$_{\odot}$ per year (Jackson et al. 1993). Such an enormous rate cannot reach the MBH, unless it is much more luminous than observed. On the other hand, it has been shown that the presence of an optically thick disc in the inner 0.03 parsec can be excluded by the absence of eclipses of the star S2 and by the absence of IR radiation due to reprocessing of the stellar flux (Cuadra et al. 2003). So the major fraction of the inflow must be hiding somewhere, perhaps in very dense molecular clumps too small to be Jeans unstable, which circularize at a short distance from the MBH, waiting for the accumulation of enough material to trigger a new episode of accretion and star formation onto the MBH.  Note that the reservoir constituted by the CND is emptied in at least 3 10$^6$ years. 

 It is difficult to deduce the present accretion rate of the MBH precisely (cf. Quataert et al. 1999; Quataert \& Gruzinov 2000b; Bower et al. 2005), but it appears that this rate can be provided by the hot stars, with no need of an accretion disc at the parsec scale (Cuadra et al. 2005, 2006). 
That these hot stars formed 5 million years ago and that the accretion disc at the parsec scale is presently absent prove that the disc has not lived continuously for {\it more} than one or two million years, and that no further star formation episode took place. 	
	But what was the {\it minimum duration} of this episode?  Had it exhausted all the gas of the disc immediately?  Until now this question has not been resolved. On simple probability considerations, we can deduce that such star formation episodes should be relatively frequent in the life history of quiescent galaxies. Indeed young massive stars whose kinematics is consistent with a circular stellar disc in Keplerian rotation around the supermassive black hole are now observed also in the inner parsec of the Andromeda nebula (Bender et al. 2005). This is problably the first of a series, and similar stellar rings will be discovered in the future in other quiescent galaxies with the development of optical interferometric techniques, showing that a large fraction of quiescent galaxies possesses such rings of massive stars, implying frequent starburst events very close to the MBH.
		
	On the other hand  low luminosity AGN (LLAGN, 10$^{41-42}$ erg s$^{-1}$) are observed in a large proportion of galaxies (Ho, Filippenko \& Sargent 1997), implying that accretion episodes occupy about 10$\%$ of the lifetime of quiescent galaxies.  Nayakshin (2004) suggests that LLAGN go through episodes of quiescence with outbursts of accretion and star formation. It is thus tempting to identify these LLAGN  accretion events with the star formation episodes that occurred in the GC and in M31. Since they took place a few million years ago, their total duration should be at least of one million years. And since many short episodes would contradict the well-defined age of the hot stars in the GC, we deduce that the episodes should have been continuous for one million years. It means that the discs should have survived during a much longer time than the formation time of the massive stars (about 10$^4$ years, cf. Nayakshin 2006, and below), so the stars should have formed at a low rate, or at least they should not have prevented subsequent accretion.  
	
	 Two different kinds of models have been proposed to account for the accretion/star formation episode in Sgr A*.
In the first one, a gravitationally unstable stationary disc is supplied {\it continuously} at a high and constant rate at least on a viscous timescale. The flow is self-regulated by star formation and  leads to a starburst in the region located at 0.1 pc, while allowing a small proportion of gas to reach the inner regions and be accreted by the black hole (Thompson et al. 2005). This model is only marginally consistent with the maximum duration of the star formation episode, which is shorter than the viscous time scale ($\ge 3\ 10^6$ years).

In the second model, a rapid episode of accretion and star formation took place, leaving the disc exhausted of its gas, through star formation and/or through accretion by the MBH, after a time much shorter than the viscous time of the disc.	
	Nayakshin (2006) proposes that the lifetime of the gaseous disc in the GC could have been prolonged by the fact that, once the first stars are formed, the heating they provide is enough to maintain the disc in a stable state during a shorter time than the viscous time. In this model, the accretion rate is not constant across the disc, and the disc is in dynamical and thermal, but not in viscous, equilibrium. If it would last longer, it would reach a viscous equilibrium where its radial structure would be reorganized, and additional angular momentum would be necessary in order to keep the disc in this stable, or marginally stable, state. This is compatible with the duration of the episode.
However the initial conditions require a delicate fine tuning so all locations of the disc reach marginal stability at the same time. The stars must begin to form at a high rate immediately after the disc is settled. On the other hand, they must not form too rapidly during the formation of the disc itself, or the gas would be destroyed. The formation of a disc able to form stars is a relatively long process, as it should satisfy two requirements:
	
\noindent - First, a geometrically thin disc. If the disc is for instance created by two molecular clouds colliding and losing their angular momentum, they should also lose their quasi-spherical structures, and the time it would take to become a thin disc in hydrostatic equilibrium with the MBH is on the order of their dimension divided by the sound velocity, i.e. about 10$^6$ years.

\noindent - Second, a Keplerian rotation. This would also take about this much time.

\noindent From this, the process is compatible with an episode lasting about 10$^6$ years.

	It is interesting to note that the amount of stellar heating  needed to quench subsequent star formation in Nayakshin's model is $\ge$ 10$^{40}$ ergs s$^{-1}$, according to his computation where the stellar mass is assumed to be 10$^{-3}$ the disc mass, and the stars are assumed to have a mass of  0.1 M$_{\odot}$ and to be accreting at their Eddington rate. The stellar luminosity is about the same as the black hole accretion luminosity, if at this epoch it was accreting at 0.01 of its Eddington rate. Extending this result to LLAGN means that the accretion and the stellar luminosity should be similar, as it indeed seems to be observed. 
	
	Note that during the present ``quiescent" phase of Sgr A*, there could also be some small intermittent accretion episodes, as proved by the discovery by Revnivtsev et al. (2004) of a hard X-ray emission of Sgr B2. It was presumably emitted 300 to 400 years ago by Sgr A* and Compton-scattered on Sgr B2. This leads to concluding that Sgr A* was at this time a kind of ``very low luminosity AGN" (10$^{39}$ erg s$^{-1}$). But in such episodes the gas should not transit through an accretion disc on the parsec scale (or star formation would take place), and regions closer to the MBH should  pick up the required material directly, most likely devoid of angular momentum.
		
	We conclude from this discussion that though there is presently no massive accretion disc around Sgr A*, there should have been one a few million years ago, and it should have lasted for about one million years  {\it  if such starburst episodes are identified with LLAGNs}.  Star formation could have taken place early  and could have been inhibited later by stellar heating during the rest of the disc lifetime, where the luminosity was dominated by newly formed stars. These  episodes are expected to occur many times during the lifetime of a galaxy.  But the problem of the initial formation of the accretion disc is still open. 
Note that they should lead to a large population of stellar remnants  (neutron stars and black holes) and of solar mass stars in the inner parsec around the MBH. If these episodes occur every 10$^7$ years,  it would mean that there should be presently about 10$^5$ stellar remnants in the inner parsec of the GC, which is indeed compatible with the observational constraints (Deegan \& Nayakshin 2007). 
		
\subsection{The accretion /  star formation scenario in luminous AGN}

The accretion process in quasars and in luminous Seyfert nuclei is probably quite different from that in the LLAGN: it is due to mergers and close interactions of the host galaxy with other galaxies. This raises two questions: is the accretion continuous or intermittent like in the GC? Does a disc exist still at the parsec scale? 

 Concerning the first question, one can easily imagine that the accretion is continuous and even close to the Eddington rate during a time comparable to the interaction (10$^8$ years), which is also the growing time of MBHs. 
There are at least two arguments in favor of this possibility:

\noindent 1. the presence of very massive BHs ($M \ge 10^9$ M$_{\odot}$) in  high redshift ($z \sim 6$) quasars, implying a growth time of a few hundred million years, which would not leave any room for intermittent accretion; but one could argue that in these objects the accretion process is different from AGN and less distant quasars;

\noindent 2. the simultaneous existence of extended radio lobes at the million parsec scale and of compact radio jets in the same direction at the parsec scale, in radio quasars and in FRI and FRII galaxies\footnote{The latter are assumed to be radio quasars seen edge on, according the the Unified Scheme, cf. Barthel 1989; this is confirmed by the presence of several AGN signatures, like broad infrared lines.}. This requires a causal link between the two structures, implying a continuous activity during at least 10$^7$ years. 

Concerning the second question, we know that QSO and AGN accretion discs are gravitationally unstable beyond 1000$R_{\rm Schw}$  (0.01 pc for a MBH of 10$^8$ M$_{\odot}$). 
Since a mechanism is required to transfer mass from large distances at a high rate, the accretion disc is certainly not limited to a dimension $\le$ 0.01 pc, unless low angular momentum gas is provided in the form of a quasi spherical accretion flow.
		
	Several observations seem to prove the presence of dense gas in Keplerian motion up to much more than 10$^3$ $R_{\rm Schw}$. Maser spots in NGC 4258 (Greenhill et al. 1995) are observed at a few tenths of a parsec (corresponding to 10$^5$ $R_{\rm Schw}$) in perfect Keplerian rotation. The broad Balmer lines in Seyfert galaxies are emitted by a gas located at about 10$^4$ $R_{\rm Schw}$ (the broad line region or BLR), and a part of the BLR is probably a flat medium illuminated by the central continuum and rotating with a nearly Keplerian velocity (in NGC 5548, Peterson \& Wandel 2000; in Mrk 110, Kollatschny 2003), which could be made of clouds ejected from the disc (Murray \& Chiang 1997). If the near-infrared continuum in quasars and AGN is produced by the accretion disc, like the visible and UV continuum, as confirmed by the continuity in time lags and size of the emission region (in Fairall 9, for instance, see Clavel et al. 1989), its radius should be  larger than 10$^3$ $R_{\rm Schw}$ (see below). All these results argue in favor of large accretion discs, though they do not imply that the discs are homogeneous or stationary. Note, however, that the rapid flux variations observed in quasars are not due to changes in the accretion rate in the outer disc, but to thermal or dynamical instabilities, or to sporadic changes in the accretion rate, taking place in the inner disc at a few tens of $R_{\rm Schw}$.
	
	In the present paper, we thus assume that the gas arriving in the nucleus has a relatively large angular momentum and circularizes at a radius larger than 0.01 parsec.
We thus first study a {\it stationary disc}, i.e. supplied permanently during at least 10$^7$ years at a rate corresponding to the luminosity of the MBH. In this case one should seek a mechanism able to avoid the destruction of the disc through star formation. It would occur if the disc is prone to only a modest star formation rate during this time, being for instance marginally stable or made of gravitationally stable clumps. Second, we examine the possibility that the disc,  like the GC,  is {\it not fed permanently}, and we discuss what consequences it would have on star formation and on the feeding of the black hole.
	
\section{Marginally stable stationary discs: theoretical considerations}	

A geometrically thin accretion disc becomes gravitationally unstable when the Toomre parameter $Q$ is smaller than unity (Toomre 1964; Goldreich \&
Lynden-Bell 1965). It is defined  as
\begin{equation}
Q = {\Omega c_{\rm s}\over \pi G\Sigma}\ ={2\sqrt{1+\zeta}\over \zeta},
\label{eq-Toomre}
\end{equation}
where $\Omega$ is the angular velocity, $\Sigma$ the 
surface density $2\rho H$, $\rho$ being the density and $H$ the scale height. In the second equation we have expressed $Q$ in terms of  $\zeta=4\pi G \rho/\Omega^2$, which measures the self-gravity\footnote {$\zeta$ is generally neglected with respect to unity, although it can be very large for $Q$ smaller than unity.}.  

When $Q$ is close to unity, the instability growth rate becomes very low, and, according for instance to Wang \& Silk (1994) the time for star formation $t_{\rm form}$ is 
\begin{equation}
t_{\rm form} = \Omega^{-1} {Q\over\sqrt{1-Q^2}}.
\label{eq-tform}
\end{equation}
It is thus necessary for a stationary disc to be marginally stable ($Q\sim 1$) to avoid rapid transformation into stars or compact bodies.
	
Goodman (2003) argues that, in order to feed AGN at rates close to the Eddington accretion rate, while maintaining the disc in a marginally stable state,
the required level of energy feedback from star formation should be impossibly large.  This argument was repeated in several papers and was used to dismiss a continuous episode of star formation in accretion discs. 

The essence of the problem lies in the amount of momentum transport that the disc can sustain. The gas in a marginally stable disc must be transported at a relatively high velocity to maintain the required accretion rate, or it will be strongly heated. Goodman (2003) discusses several ways of providing the disc with an enhanced momentum transport, and concludes that they were all difficult to achieve, so the transport had to rely on standard turbulent viscosity. Consequently, Sirko \& Goodman (2003) studied the case of discs with standard viscosity and show that any heating mechanism able to maintain the disc in a marginally stable state would lead to the a strong infrared emission that contradicts the observed
spectrum of a typical AGN.

In CZ99, it was shown that supernovae can ensure angular momentum transport, while the newly formed stars provide modest complementary heating, both being sufficient for maintaining the disc in a marginally stable state.
 On the contrary, in the present paper we will try to account for the observed spectrum, without imposing a rate of star formation inside the disc that is consistent with the rate of supernovae necessary to transport the angular momentum. 

Let us recall the problem here. 
The standard stationary ``$\alpha$-disc" in its simplified vertically-averaged form (Shakura \& Sunyaev 1973) is governed by a set of algebraic equations allowing solution of the radial structure as a function of only 3 parameters: the mass of the MBH, $M$, the accretion rate, $\dot{M}$, and $\alpha$, the ``viscosity parameter"\footnote{some authors, like Goodman, add another free parameter to distinguish between a viscosity based on gas pressure or on total pressure. We have adopted the total pressure prescription in our studies.}. 

The equation accounting for the transport of angular momentum is given by
\begin{equation}
\dot{M} f(R) = 6\pi \nu \rho H=6\pi\alpha \rho c_{\rm s} H^2,
\label{eq:nuviscc}
\end{equation}
where $H$ is the disc scale height, $\rho$ the vertically averaged density, and $c_{\rm
s}$ the isothermal sound speed $\sqrt{P/\rho}$, $P$ being the sum of the gas  and radiation pressure; $\nu$ is the kinematic viscosity
given by the $\alpha$-prescription, $\alpha$ being a dimensionless number $\le$ unity; $f(R)$ accounts for 
the inner boundary conditions. It is equal to unity in the outer regions of interest for us, so we will ignore it in the following. 

The accretion rate also intervenes in the energy balance between local viscous dissipation and
radiative cooling:
\begin{equation}
F_{\rm rad} = \sigma T_{\rm eff}^4=\frac{3 \Omega^2 \dot{M}}{8 \pi},
\label{eq:ebb}
\end{equation}
where $\Omega$ is the Keplerian angular velocity, $\sigma$ the Stefan constant, and $F_{\rm rad}$ the flux radiated by each face of the disc. 

The inner part of the accretion disc is assumed to be a standard $\alpha$-disc, like in Sirko \& Goodman.  When $Q$ reaches unity, it becomes marginally stable, and the prescription $Q=1$  is imposed  (hereafter we call it a ``Q-disc").
Thus in the Q-disc there is the additional constraint $Q=1$ (corresponding to $\zeta=2(1+\sqrt{2})$=4.8) with respect to the $\alpha$-disc. It leads to a given dependence of the density on the radius: $ \rho= \sqrt{1+\zeta}\ \Omega^2 / (2 \pi G) \propto R^{-3}$. One must therefore suppress one of the equations in the set governing the $\alpha$-disc. 

Here lies the fundamental difference between us and Sirko \& Goodman. They chose to suppress the energy equation \ref{eq:ebb}, assuming that ``some feed-back mechanism supplies just enough additional energy to the disc to prevent $Q$ falling below the minimum $Q\sim 1$." And they kept the momentum equation \ref{eq:nuviscc}. In contrast, CZ99, following Collin \& Hur\'e (1999),  kept the energy equation \ref{eq:ebb}, but suppressed  Eq. \ref{eq:nuviscc}, assuming that some additional momentum transport mechanism exists in the disc and maintains it in a marginal state. In CZ99, this momentum transport was identified with the effect of the expansion of supernovae remnants in the disc, and the contribution of the star heating to the energy equation was taken into account.

Why was the amount of extra heating so important in the Sirko \& Goodman paper? Simply because, as they assume a  turbulent viscosity in the Q-disc and moreover chose a low value for the $\alpha$-parameter ($\alpha$=0.01), the transport was very slow, and consequently the disc surface density was high, requiring strong heating. In CZ99, the required heating was much lower, as the transport mechanism was more efficient than a purely viscous one. Less gas - therefore less extra heating - was necessary.  The stellar heating was locally stronger than the gravitational heating, but it was negligible compared to the whole disc emission.  

\subsection{Mass transport}

Several causes can compete in transporting angular momentum - supernovae, disc-driven winds or jets, stars captured by the disc, cloud collisions, etc. We focus here on the transport of angular momentum related specifically to the star formation process, i.e. supernovae explosions. Since we assume that the disc is living for more than 10$^7$years, massive stars have indeed amply time to evolve.
This mechanism was proposed first by Rozyczka et al. (1995). The idea is that, when a supernova explodes in the accretion disc and the shell expands, a fraction of the total momentum ($P_{\rm tot}$, about 10$^{43}$ g cm s$^{-1}$ for a supernova producing a kinetic energy of  10$^{51}$ ergs) delivered by the supernova is supplied to the disc. Their 2-dimensional simulations of a supernova exploding in a thin Keplerian accretion disc have shown that it leads to a redistribution of the angular momentum.   This is because the leading 
hemisphere of the supernova
shell has an excess of angular momentum compared to the disk, while  
the trailing  hemisphere has
a deficit of angular momentum. Consequently the  leading 
hemisphere is driven towards the center and the trailing one towards 
the exterior. 
Rozyczka et al.  made an analytical estimate of the angular momentum given to the disc, which agreed with their 2-D simulations. In CZ99 we questioned their estimation and proposed another one, which unfortunately lowers the efficiency of the process (see Appendix).  Clearly, 3-dimension simulations are needed to settle the question; in the meantime we shall consider both prescriptions. 

In the following we divide the momentum transport into two parts, one provided by the standard turbulent viscosity (Eq. \ref{eq:nuviscc}), $\dot{M}_{\rm viscous}$, and one given by the additional mechanism, $\dot{M}_{\rm non\ viscous}$. 

\subsection{Non viscous heating}

The shape of the SED in the near infrared band has always been a problem, as it is not well-fitted by the standard disc emission (see for instance Koratkar \& Blaes 1999 for a review). It is clear that, besides viscous heating, another process contributes to the heating in the outer regions.  This extra heating can have several causes: gravitational instability itself (Lodato \& Bertin 2001), illumination by the central source (for instance if the disc is warped), to illumination by stars embedded in the disc or outside, etc.
 We do not distinguish between these different processes, because they can have the same parametrization in terms of an effective temperature as a function of the radius.   If we hypothesize that it is due to stars, it will give us {\it the maximum} number of stars needed to achieve the heating. Note that the heating provided by stars outside the disc is a factor two smaller than if they were trapped within the disc. Even if the disc is not thick in the sense of the Rosseland mean, a surface density as low as 10$^{-3}$g cm$^{-2}$ (which is never reached in this problem) is indeed amply sufficient for absorbing the UV radiation coming from external stars in a thin layer at the disc surface, and for reprocessing it as infrared radiation towards the interior of the disc (cf. Collin \& Hur\'e 1999 for a discussion of this effect). The only difference with stars embedded in the disc is that half of the reprocessed radiation is emitted towards the exterior. The number of stars is thus estimated simply by assuming that $L_{\rm non\ viscous}$ (2$L_{\rm non\ viscous}$ if the stars are not embedded in the disc) is due to hot massive stars radiating 10$^{38}$ erg s$^{-1}$ each\footnote{In CZ99, the question was treated in a slightly more detailed way as we considered the optically thin and the optically thick cases separately, and we assumed that the stars were embedded in the accretion disc.}. In Nayakshin (2006), the stellar heating was ascribed to low-mass stars (0.1 M$_{\odot}$) in an Eddington-limited process of accretion; but this process lasts only 10$^4$ years, and thereafter, all stars are transformed into massive stars, so it is not valid for a stationary long-lasting disc.

 The non-viscous heating is taken simply as $L_{\rm non\ viscous}=\pi R^2 F_{\rm non\ viscous}$, where  $F_{\rm non\ viscous}$ is
\begin{equation}
F_{\rm non\ viscous} = \sigma T_{\rm eff}^4\ -\ \frac{3 \Omega^2 \dot{M}_{\rm viscous}}{8 \pi}.
\label{eq-Fstar}
\end{equation}

\subsection{Time scales}

Two time scales are of interest for this scenario. One is the viscous time, $t_{\rm visc}$, equal to $M_{\rm disc}/\dot{M}$ where $\dot{M}$ is given by Eq. \ref{eq:nuviscc}, and $M_{\rm disc}$ is the disc mass, here computed simply as $M_{\rm disc}=\pi R^2 \Sigma$. The second is the time $t_{\rm accr}$ it takes for the stars to acquire a high mass by accretion. 

Once a clump has collapsed (in a dynamical time of $\sim 1/\Omega$) and a protostar is formed, it undergoes
a mechanism of accretion
and growth, such as proposed by Artymowicz, Lin \& Wampler (1993) for stars trapped by the accretion disc.  A star of mass $m_*$ at a radius $R$ can accrete all the gas contained in its Roche lobe of radius
\begin{equation}
R_{\rm H}=\left({m_*\over M}\right)^{1/3}R.
\label{eq-roche}
\end{equation}
The accretion rate is given by the shear velocity corresponding to the radius of the Roche lobe at each moment $\sim \Omega R_{\rm H}$ and is thus
\begin{equation}
\dot{m_*}\sim \Omega R_{\rm H}^2 \rho H .
\label{eq-maccr1}
\end{equation}
One finds the accretion rate,
\begin{equation}
{\dot{m_*} \over M} ={1\over t_{\rm accr}} \left({m_*\over M}\right)^{2/3},
\label{eq-maccr2}
\end{equation}
where $t_{\rm accr}=M/(\rho H R^2 \Omega)$, and finally at a time $t$
\begin{equation}
{m_* \over M} - {m_{\rm init}\over M} =\left({t\over 3t_{\rm accr}}\right)^3,
\label{eq-maccr3}
\end{equation}
where $m_{\rm init}$ is the initial mass of the star. 

However, we will not assume that the star can reach its Roche mass, which is very high (up to 10$^5$M$_{\odot}$!), but that the accretion stops when it approaches the outflowing wind rate, which probably occurs when the mass is close to a few tens of M$_{\odot}$. The accretion rate is also limited by the Eddington rate. We call $t_{\rm accr}$ the time it takes to acquire a mass of 10M$_{\odot}$. 

\subsection{Opacity}

Opacity plays a fundamental role in this problem, as it relates the mid-plane temperature $T$ of the disc to the effective temperature $T_{\rm eff}$ through the
 diffusion equation, which can be written as
\begin{equation}
\sigma T_{\rm eff}^4={\sigma T^4\over 3\tau/8+1/2+1/4\tau},
\label{eq:Fradbis}
\end{equation}
where $\tau$ is the frequency averaged opacity
\begin{equation}
\tau=\rho H \kappa ={\Sigma \kappa \over 2},
\label{eq:taubis}
\end{equation}
and $\kappa $ the opacity coefficient in g$^{-1}$ cm$^2$.
 Since $T$ determines the radiation and gas pressures (cf. the appendix), and therefore the scale height of the disc, the whole disc structure depends on the opacity. In CZ99 we used a scaling law, assuming that $\kappa $ was a constant on the order of unity, which is only valid for $T\le 100$K. In fact $\kappa $ varies considerably in the range of temperature of interest to us in this paper, i.e. around 10$^3$K. 

Actually, one should make a distinction between the Rosseland and the Planck frequency-averaged opacities, the first being valid in the optically thick case, the second in the optically thin one. In Collin \& Hur\'e (1999), the distinction was performed according to the computations of Hur\'e et al. (1994). Here we decided not to use the Planck opacities, which would have complicated the problem unnecessarily, as there is almost always an optically thick solution in the regions of interest to us. We use interpolations of the mean opacities computed  by Ferguson et al. (2005), which cover the whole range of temperature and density needed without any extrapolation.

\section{Marginally stable stationary discs: results}

We refer to CZ99 for discussions concerning the survival of the accretion disc to the perturbations caused by the stellar formation and the presence of hot massive stars in the disc or by the supernovae explosions.

We computed 3 models. The first one corresponds to a luminous quasar (mass equal to 10$^9$ M$_{\odot}$, Eddington ratio equal to 0.5), the second to a luminous Seyfert nucleus  (mass equal to 10$^8$ M$_{\odot}$, Eddington ratio equal to 0.1), and the third to a weak Seyfert nucleus (mass equal to 10$^7$ M$_{\odot}$, Eddington ratio equal to 0.01). The bolometric luminosities and the accretion rates are, respectively, 

\begin{itemize}

\item for the quasar: $L$= 0.75 10$^{47}$ erg s$^{-1}$, accretion rate = 23 M$_{\odot}$ yr$^{-1}$, 
\item for the luminous Seyfert: $L$= 1.5 10$^{45}$ erg s$^{-1}$, accretion rate =  0.46 M$_{\odot}$ yr$^{-1}$,
\item for the weak Seyfert: $L$=  1.5 10$^{43}$ erg s$^{-1}$, accretion rate = 4.6 10$^{-3}$ M$_{\odot}$ yr$^{-1}$.

\end{itemize}

Since one aim of this paper is to show that the SED emitted by the disc heated by a non viscous process can agree with AGN observations, we first determine the rate of heating (i.e. $\sigma T_{\rm eff}^4$) as a function of the radius, so as to obtain a spectrum in agreement with a typical AGN SED (the one given by Sirko \& Goodman 2003). The theoretical  SED is computed by assuming a local black body emission at the effective temperature $T_{\rm eff}$:
\begin{equation}
\nu L_{\nu}= {16\pi^2 cos(i) h\nu^4\over c^2}
\int_{R_{\rm in}}^{R_{\rm out}} {RdR \over exp(h\nu/kT_{\rm eff})-1},
\label{eq-lum-em}
\end{equation}
where $cos(i)$ is the inclination of the disc axis on the line of sight (in principle $i \le \pi /4$ for a type 1 AGN according to the Unified Scheme), and ${R_{\rm in}}$ and ${R_{\rm out}}$ are respectively the inner disc radius (here 3 $R_{\rm Schw}$) and the outer radius (here 10$^6$ $R_{\rm Schw}$). The factor  $16\pi^2 $ accounts for the fact that the luminosity of an AGN is deduced from the flux observed at Earth, assuming that it is a point source radiating isotropically. We take an inclination $cos(i)=0.75$. We assume a Schwarzschild MBH with an efficiency $\eta=0.057$, and a viscosity coefficient $\alpha=0.3$ (this parameter is necessary not only for the $\alpha$-disc, but also for the Q-disc to compute the viscous contribution to the heating and to the momentum transport). A good fit to the spectrum in the visible and infrared bands is obtained for  $T_{\rm eff}\propto R^{-\gamma}$ with  $\gamma$=0.47, starting at a given radius, which we call the transition radius, $R_{\rm trans}$. In the quasar case, the radius where the disc becomes gravitationally unstable, $R_{\rm marg}$, being small (300 $R_{\rm Sch}$),  $R_{\rm trans}$ is chosen equal to $R_{\rm marg}$.  In the other cases, $R_{\rm trans}$ is smaller than $R_{\rm marg}$.  It is indeed logical that the heating due to stars begins to be smoothly efficient before $R_{\rm marg}$, since the stars can surround the disc or might have migrated towards the MBH (cf. CZ99). Note that the value of $\gamma$ is close to 0.5, which corresponds to a uniformly distributed source of heating when it dominates\footnote{For an illumination by the central source, $\gamma$= 0.5 can be due to a strong flaring or warping of the disc, or to a hot medium back-scattering the radiation of the central source (cf. Collin \& Hur\'e 1999).}.

 The method to compute the SED is oversimplified, but it has the advantage of being relatively correct  for an optically  thick medium in the visible and infrared ranges where the Q-disc is emitting. Indeed  the critical wavelength $\lambda_{\rm crit}$ corresponding to the beginning of the emission of the Q-disc is close to $hc/(4kT_{\rm eff}(R_{\rm marg}))$.   $T_{\rm eff}(R_{\rm marg})$ is $\sim  3000$K. Thus $\lambda_{\rm crit}\sim 10000$ ${\rm Angstr\ddot{o}}$ms. This method is very bad for the UV emission produced by the $\alpha$-disc, due to electron scattering opacity that distorts  the spectrum towards high energy and makes it ``flatter". In the X-ray range, the emission does not stem from the accretion disc, but most probably by a hot medium lying above the disc, like a corona or distributed magnetic flares (Haardt \& Maraschi 1991, 1993, and many subsequent papers). 

\begin{figure}
\begin{center}
\includegraphics[width=7cm]{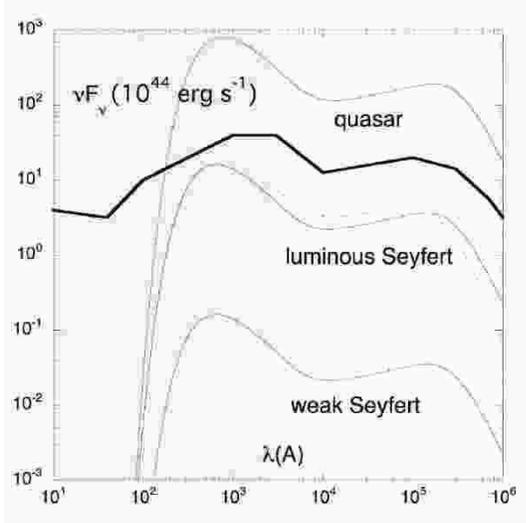}
\caption{The computed SED for three typical models of  a quasar and a Seyfert nucleus,  corresponding to Figs. 2, 3, and 4, compared to a typical AGN SED. In all cases the internal and the external radii of the discs are  equal to 3 and 10$^6$ $R_{\rm Schw}$, respectively. The dotted line corresponds to the luminous Seyfert with a clumpy disc described in the second part of the paper. The grey zone shows the region where the SED is not computed correctly and is flatter than displayed here. }     
\label{spectres}
\end{center}
\end{figure}

Figure \ref{spectres} displays the three theoretical SED and the observed AGN spectrum.   They agree reasonably well  within the range of wavelengths allowed by the approximation of Eq. \ref{eq-lum-em} (i.e. above 3000 ${\rm Angstr\ddot{o}}$ms), taking into account the differences of SED from one object to the other and the difficulty of defining a ``typical" AGN spectrum.

\begin{figure*}
\begin{center}
\includegraphics[width=13.5cm]{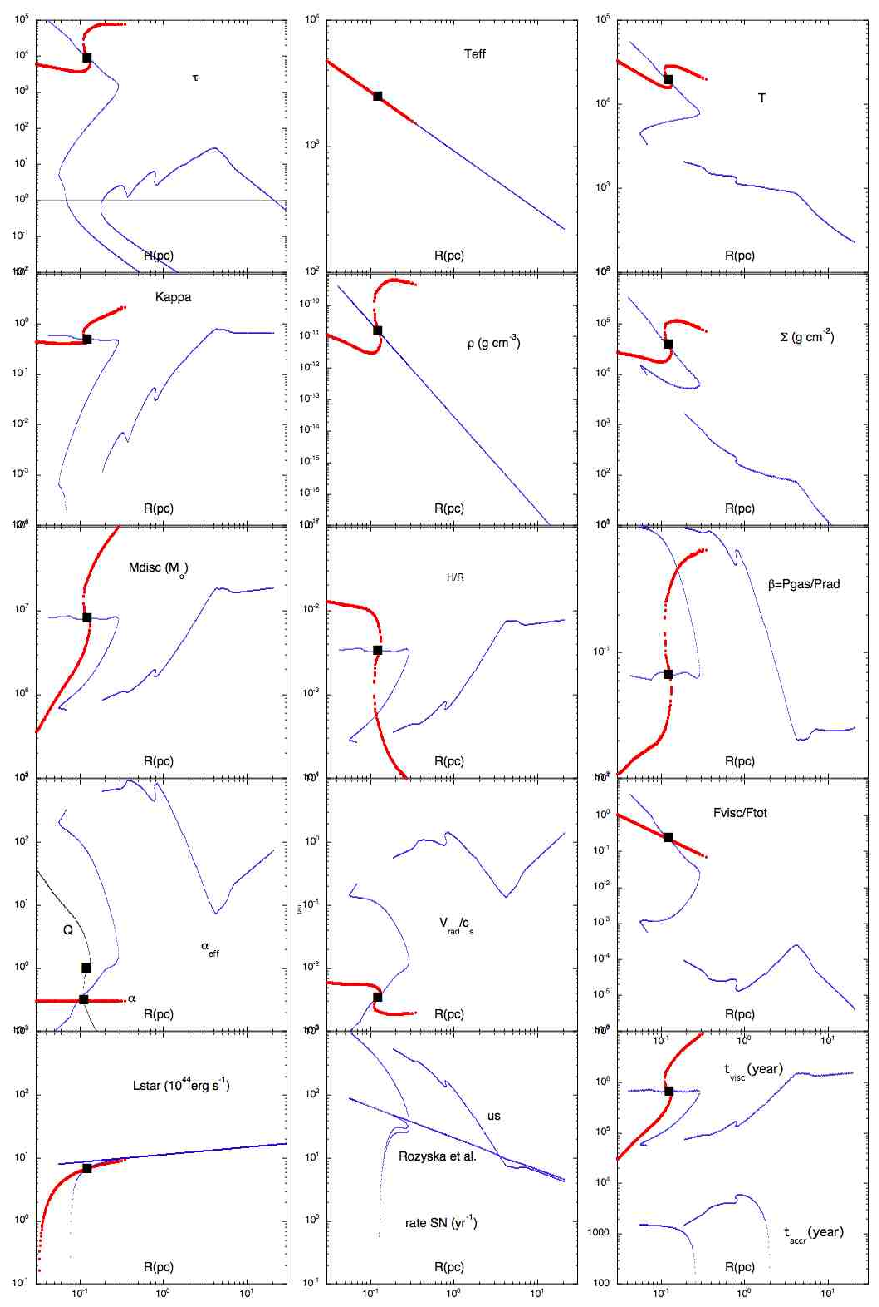}
\caption{Dependence of several parameters on the radius, for the model corresponding to the quasar spectrum shown in Fig. \ref{spectres}. The parameters of the model are  $M=10^9$ M$_{\odot}$,  Eddington ratio 0.5, $\alpha=0.3$, and a dependency of $T_{\rm eff}$ versus $R$:  $T_{\rm eff}\propto R^{-0.47}$, with $R_{\rm trans}$ equal to 0.03 pc (300 $R_{\rm Schw}$, about $R_{\rm marg}$). In the first figure (left and top panel), the optical thickness is displayed for all solutions, but the optically thin solutions have been suppressed in the other figures. The $\alpha$-disc parameters are shown with big red dots, and the Q-disc with small blue dots. The junction between the $\alpha$ and the Q discs is shown by the big dark square.}     
\label{solutions-marg1} 
\end{center}
\end{figure*}

The results corresponding to the 3 typical cases considered here are displayed in Figs. \ref{solutions-marg1}, \ref{solutions-marg2}, and \ref{solutions-marg3}, and the model equations are detailed in the Appendix. These figures show the dependence of several parameters on the radius: the optical thickness $\tau$, the effective temperature $T_{\rm eff}$, the mid-plane temperature $T$, the opacity $\kappa$, the density $\rho$, the surface density $\Sigma$, the disc mass $M_{\rm disc}=\pi \Sigma R^2$, the aspect ratio $H/R$, and the ratio $\beta$ of the gas pressure $P_{\rm gas}$ to the total pressure $P_{\rm gas}+P_{\rm rad}$. The following figure (left of the fourth row) shows the value of $Q$ in the $\alpha$-disc, and the value of $\alpha_{\rm eff}$ in the Q-disc, allowing a check that the junction between the two discs occurs when $\alpha_{\rm eff}$ (an ``effective" viscosity parameter given by Eq. \ref{eq:nuviscc}) is equal to 0.3, the value of $\alpha$ in the $\alpha$-disc, and when $Q=1$ in the $\alpha$-disc (except for the weak Seyfert case when we did not find any solution for the $\alpha$-disc below $Q=3$). Next are displayed the ratio of the radial to the sound velocity $V_{\rm rad}/c_{\rm s}$, the ratio of the viscous to the total flux, the stellar luminosity $L_{\rm star}$ (assuming that the additional heating is due to stars), the rate of supernovae explosions required to transport the angular momentum, and finally the viscous time $t_{\rm visc}$ and the accretion time $t_{\rm accr}$ in the Q-disc.

Because of the additional heating, $Q$ is increased in the $\alpha$-disc that extends farther than in the absence of this heating. Of course this makes the implicit assumption that, when the disc is heated, it adjusts its structure to the new conditions, which means that its lifetime is longer than the viscous time ($\sim 10^6$ years for the quasar and the luminous Seyfert, $\sim 10^7$ years for the weak Seyfert, see Figs. \ref{solutions-marg1}, \ref{solutions-marg2}, and \ref{solutions-marg3}). 

An immediately striking result is that there are several solutions at the same radius  in some regions, corresponding to an optically thick and an optically thin disc. This also occurs in the $\alpha$-disc (cf. Fig. 3). Except in the first panel (left of the first row), we only show  those solutions corresponding to an optically thick disc, since it is the only one to give the spectrum computed with Eq. \ref{eq-lum-em}.
Even the optically thick solution is not unique and there are sometimes three optically thick solutions at the same radius, two with the temperature decreasing with the increasing radius being thermally stable, and the intermediate one thermally unstable. Note that the junction between the $\alpha$-disc and the Q-disc corresponds to the stable upper solution.  

This phenomenon is well-known, for instance, in the case of dwarf novae. However, here it is not due to the ``ionization instability" around 10,000K, but to dust sublimation around 1,000K. The reason of such behavior is  that the opacity $\kappa$ passes by a strong minimum (varying by many orders of magnitudes)  around 1,000K with respect to its value at 100K, where it is dominated by dust, and 10,000K, where it is dominated by bound-bound and bound-free atomic processes. When the optical thickness $\tau$ decreases, $T$ (roughly equal to $\tau^{1/4}T_{\rm eff}$) also decreases. The radiation pressure (proportional to $\tau T^4$) and the gas pressure (proportional to $T$) also undergo a big decrease, leading to the decrease in the sound velocity, therefore in the scale height and in the surface density. And finally $\alpha_{\rm eff}$ (inversely proportional to $c_{\rm s}^2 \Sigma$) can increase by 3 orders of magnitude!  Thus   $\kappa$ entirely governs the disc structure. 

Since one of the stable solutions disappears at a given radius, there should be a jump in the structure at this position. However the jump cannot be abrupt, because there will be some radial heat transport through both radiation and turbulence, which are not taken into account in our computation. It is also highly probable that a supersonic turbulence develops and contributes to the transport of momentum in this region. Moreover, the Keplerian law is no longer valid close to the transition, as there is a strong radial variation in the pressure. It is clear that our solution in this highly perturbed region is non reliable. 

 In the region where two stable solutions exist, the disc can also undergo changes in structure, passing through one solution to the other in a thermal time, which is much shorter than the viscous time. It would be interesting to see whether this thermal instability can lead to a ``limit-cycle" like in dwarf novae.  Figure \ref{fig-Teff-vs-Sig-M8} shows the behavior of $T_{\rm eff}$ versus the  surface density $\Sigma$ at different radii, for a disc with a BH mass of 10$^8$M$_{\odot}$ and an Eddington ratio varying from 10$^{-4}$ to 0.5. As expected from Figs. 3 and 4, there are multiple solutions between 0.01 and 0.04 pc. But in contrast to the dwarf novae case, there are three values of $T_{\rm eff}$ for one value of $\Sigma$, and not the opposite. The solutions with positive $\delta T_{\rm eff}/\delta \Sigma$ are stable and the one with negative $\delta T_{\rm eff}/\delta \Sigma$ is unstable. The left stable solution corresponds to low optical thickness (but greater than unity), the right one to a large optical thickness.  This situation cannot lead to a ``limit-cycle", as a stable solution exists always whatever the accretion rate (cf. Fig. \ref{fig-Teff-vs-Sig-M8}).

In all cases, the study of this difficult problem is beyond the scope of the present paper. It concerns the region located between 0.05 and 0.3 pc for the quasar, 0.01 and 0.03 pc  for the luminous Seyfert, and between 0.003 and 0.01 pc for the weak Seyfert. Though we agree that it is unsatisfactory, we leave this part of the disc out of this discussion. 

\begin{figure*}
\begin{center}
\includegraphics[width=13.5cm]{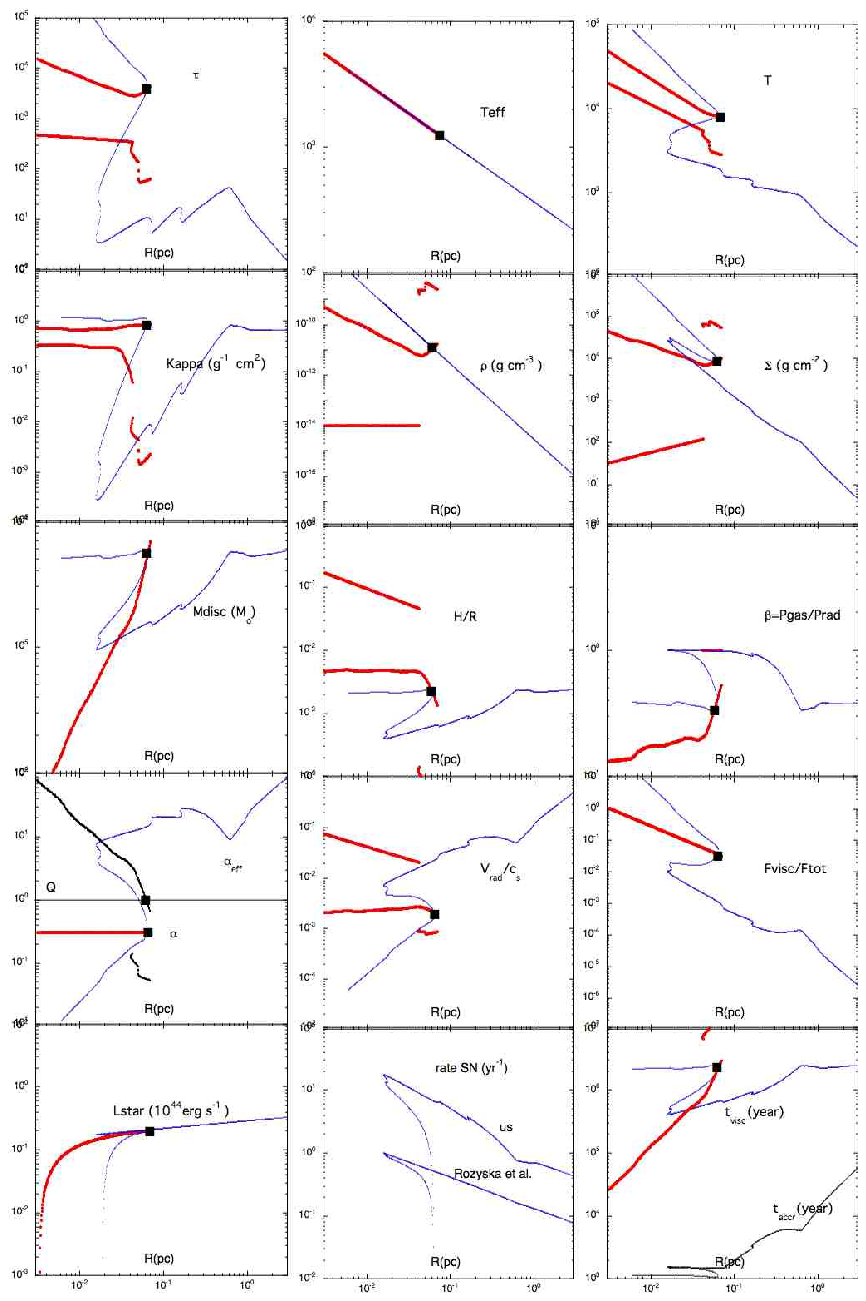}
\caption{Same as Fig. \ref{solutions-marg1}, but for the model corresponding to the luminous Seyfert spectrum shown in Fig. \ref{spectres}. The parameters of the model are  $M=10^8$  M$_{\odot}$,  Eddington ratio 0.1, $\alpha=0.3$, and a dependency of $T_{\rm eff}$ versus $R$:  $T_{\rm eff}\propto R^{-0.47}$, with $R_{\rm trans}$ equal to 0.003 pc (300 $R_{\rm Schw}$, about $R_{\rm marg}$/3). Due to the additional heating, the $\alpha$-disc is gravitationally stable up to 0.1 pc. }     
\label{solutions-marg2} 
\end{center}
\end{figure*}

\begin{figure*}
\begin{center}
\includegraphics[width=13.5cm]{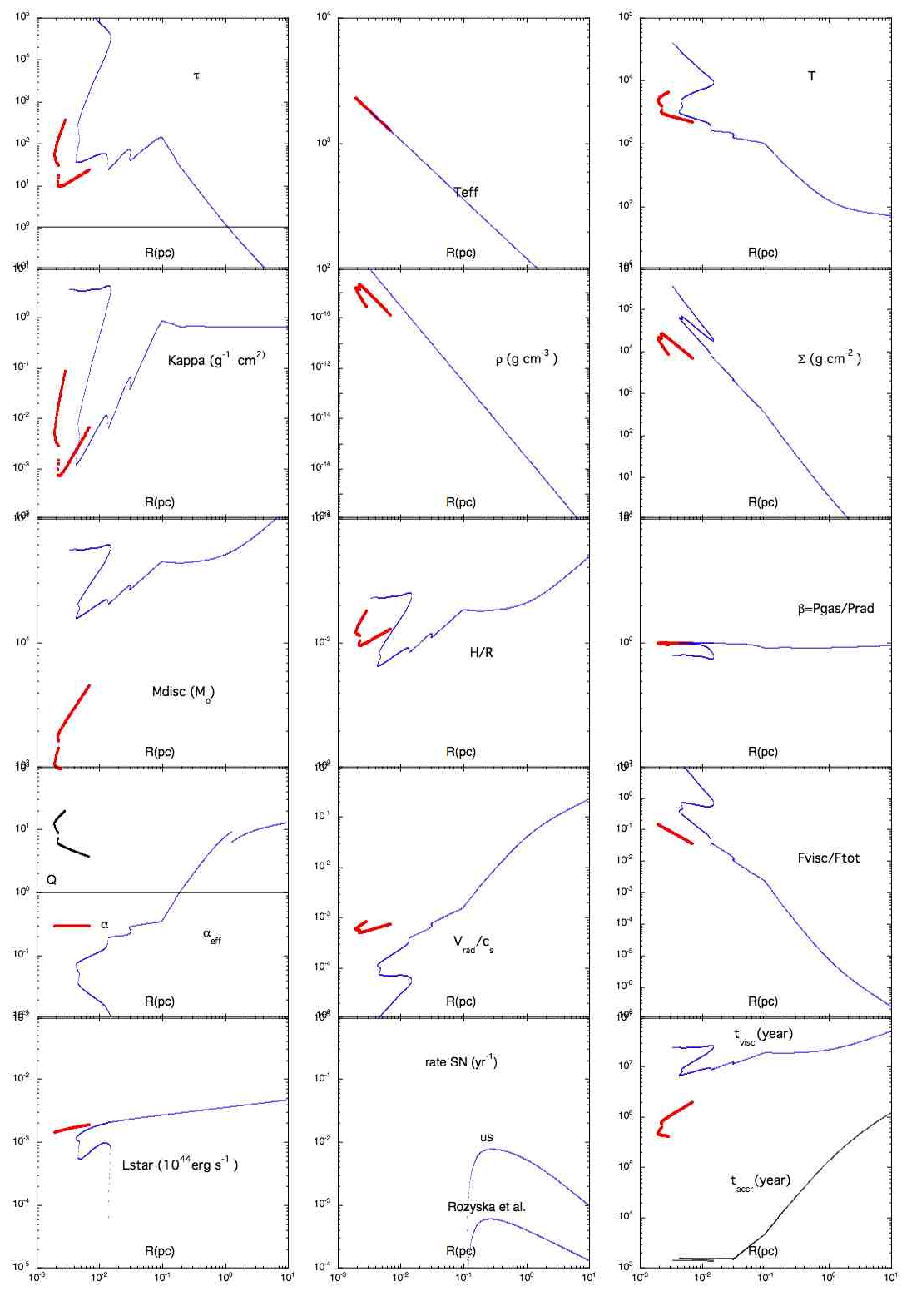}
\caption{Same as Fig. \ref{solutions-marg1}, but for the model corresponding to the weak Seyfert spectrum shown in Fig. \ref{spectres}. The parameters of the model are  $M=10^7$  M$_{\odot}$,  Eddington ratio 0.01, $\alpha=0.3$, and a dependency of $T_{\rm eff}$ versus $R$:  $T_{\rm eff}\propto R^{-0.47}$, with $R_{\rm trans}$ equal to 0.003 pc (3000 $R_{\rm Schw}$, about $R_{\rm marg}$/100). Due to the additional heating, the $\alpha$-disc is gravitationally stable up to at least 0.006 pc. }     
\label{solutions-marg3} 
\end{center}
\end{figure*}

\begin{figure}
\begin{center}
\includegraphics[width=9cm]{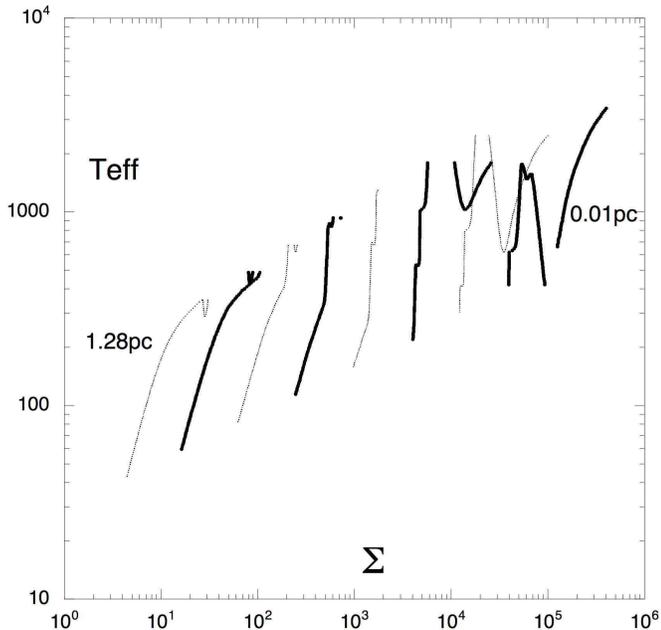}
\caption{$T_{\rm eff}$ versus $\Sigma$ for the model  $M=10^8$  M$_{\odot}$,  Eddington ratio 0.1, $\alpha=0.3$, and a dependency of $T_{\rm eff}$ versus $R$:  $T_{\rm eff}\propto R^{-0.47}$ with $R_{\rm trans}$ equal to 0.01 pc. The Eddington ratio varies from 10$^{-4}$ to 0.5 along the curves, which correspond from left to right to radii decreasing by a factor 2. To make the figure more legible, one out of two curves is in thick lines. The interruptions of the curves are due to the limitations imposed on the optical thickness and on the temperature.}     
\label{fig-Teff-vs-Sig-M8}
\end{center}
\end{figure}

Note first that {\it the Q-disc is highly perturbed due to the opacity gap within about one decade of radius, as soon as the $\alpha$-disc becomes gravitationally unstable}.
The main properties {\it concerning only the non-perturbed regions}  are:

\noindent - The mid-plane temperature, the density, and the surface density decrease with increasing radius. 

\noindent - In contrast, the mass of the Q-disc is about constant with radius and always negligible compared to that of the MBH; the disc is geometrically thin, with an aspect ratio on the order of 10$^{-3}$ to 10$^{-2}$.  

\noindent   -  $\alpha_{\rm eff}$ reaches high values, while the standard $\alpha$ cannot be greater than unity. The radial velocity varies accordingly, but it remains subsonic in the whole disc.

\noindent - We can compare the viscous time with the time it takes for a star to acquire 10 M$_{\odot}$ ($t_{\rm accr}$). The viscous time is larger than the accretion time, owing to the large rate of mass transfer, meaning that as soon as stars are formed they can be considered as massive stars.  

	Let us now discuss the most important properties related to our problem, still only in the non-perturbed regions.
	
\subsection{Stellar heating}

The non viscous luminosity is always lower by at least two orders of magnitudes than the whole AGN luminosity. 
 If we attribute this luminosity to a population of massive stars, it means that about 10$^7$ massive stars per decade of radius in the quasar case, 10$^5$ in the luminous Seyfert case, and 10$^3$ in the weak Seyfert case, are necessary at all radii. This is quite large, but not impossible if we consider that these stars can be located outside the disc.  Such a population could correspond to a scaled version of the hundred massive stars in the GC, where the accretion rate during the accretion/star formation episode was probably respectively 5, 3, and 1 orders of magnitudes lower. Since these massive stars live for about 3 10$^6$ years, during the whole active lifetime of the object (say  10$^8$ years) there should have been 30 times more massive stars and therefore stellar remnants. Thus there should be about 3 10$^8$ M$_{\odot}$ in stellar remnants around the MBH in the quasar case, 3 10$^6$ M$_{\odot}$ in the luminous Seyfert, and  3 10$^4$ M$_{\odot}$ in the weak Seyfert case. Even accounting for twice as much stellar mass due to low-mass stars (recall that the IMF is top-heavy), it stays lower than the mass of the MBH itself in the quasar case, and it is completely negligible in the Seyfert case. Note, however, that these remnants should have migrated outside the accretion disc, or the gas would be dominated by the stellar mass. Finally we recall that the heating can be provided (and is certainly partly) by another source than newly-formed massive stars, such as irradiation by the central X-ray source. Finally, in the case of an irradiation by external stars, one should take the stellar contribution amounting to  $2L_{\rm star}$ into account in the {\it observed} spectrum. This contribution should appear in the UV if it is emitted by massive stars. Though it is not negligible (it is equal to about 5$\%$ of the bolometric luminosity), it would not be as detectable as the  the disc emission.

\subsection{Rate of supernovae}

 The rate of supernovae per decade in radius and per year corresponding to our prescription and to that of Rozyska et al. (1995) decreases with increasing radius, mainly because $H$ increases (more rapidly than the radius actually), so the momentum given to the disc by one supernova increases. The rates are generally about one order of magnitude higher with our prescription than with that of Rozyska et al.. 
 
The required rates of supernovae with the Rozyska et al. prescription at the beginning of the unperturbed region are  for the quasar, the luminous Seyfert, and the weak Seyfert : 10, 0.1, and 10$^{-3}$ per year, respectively, (in the quasar case, it is the same with our prescription and that of Rozyska et al.). Since one supernova  releases a mass on the order of 10 M$_{\odot}$ in the disc as a stellar remnant or gas (of which a  fraction stays inside the disc), it corresponds to an accretion rate {\bf higher than the assumed accretion rate!} It decreases farther out, but  only becomes acceptable at 10 pc from the MBH. Only in the weak Seyfert case is it comparable to the accretion rate, and is lower than the accretion rate (which is necessary for the gaseous disc to survive) at about 1 pc. Moreover, we recall that the Rozyska et al. prescription certainly underestimates the required rate, as explained in the previous section. 

The weak Seyfert case is interesting as a supply of momentum transfer is necessary only for a radius larger than 0.1 pc. At a smaller radius, the turbulent viscosity is high enough to provide the mass transfer in the Q-disc. This is due to the additional non viscous heating, which remains compatible with the observed spectrum and corresponds to a reasonable amount of massive stars. 

	One can definitively conclude that supernovae are not able to provide the required momentum transport except in the outskirts of the Q-disc. In order for the inner accretion disc to survive star formation, it is necessary to seek another mechanism for momentum transport. Finally, the inner parts of the Q-disc are probably highly perturbed and out of equilibrium.	  
	
\section{Stationary clumpy discs}
	
\subsection{Mass transport by cloud collisions: theoretical considerations}

Duschl \& Britsch (2006) have shown that the gravitational instability itself may lead to
turbulence and thus viscosity in the disc. The other way to get momentum transport, on which we concentrate here, is that a gravitationally unstable disc should fragment into clouds whose typical  size is equal to the disc scale height. Several people have studied the ``viscosity" resulting from their interactions, and different prescriptions have been proposed (Goldreich \& Tremaine 1978; Stewart \& Kaula 1980; Lin \& Pringle 1987; Jog \& Ostriker 1988 and 1989; Shlosman \& Begelman 1989; Gammie et al. 1991; Ozernoy et al. 1998; Kumar 1999; Duschl et al. 2000; Vollmer \& Duschl 2001; Vollmer \& Beckert 2002, etc). 

We adopt a very simple approach here, assuming that the disc is made of identical clouds of dimensions $l_{\rm cl}$ that have a mean separation $d_{\rm cl}$. The basic idea is that, since the clouds have an orbital velocity depending on their distance to the MBH, they are subject to a viscous torque able to transform orbital motions into turbulent motions. They lose a fraction of their kinetic energy through physical collisions and gain kinetic energy by gravitational interactions with other clouds.  Gravitational interactions occur when the clouds approach a few times their tidal radius $d_{\rm H}=R(m_{\rm cl}/M{\rm BH})^{1/3}$ (see Eq. \ref{eq-roche} where $m_*$ is replaced by the mass of the cloud $m_{\rm cl} $). The change in velocity during these encounters is equal to the shear velocity corresponding to this radius, $\Omega  d_{\rm H}$. 

When the dimension $l_{\rm cl}$ of the clouds is close to their mean distance $d_{\rm cl}$, physical encounters occur at the same frequency as gravitational encounters. For the steady state, this leads to a dispersion velocity:
\begin{equation}
V_{\rm disp}^2\sim (Gm_{\rm cl} \Omega)^{2/3} .
\label{eq-vdisp1}
\end{equation}
When the size of the clouds is smaller than their mean distance, the mean time interval between two gravitational interactions $t_{\rm grav}$ is $\sim \Omega^{-1}(d_{\rm cl}/d_{\rm H})^2$, while the mean time interval between two physical collisions $t_{\rm coll}$ is $\sim \Omega^{-1}(d_{\rm cl}/l_{\rm cl})^2$, so the balance between losses and gains writes as
\begin{equation}
V_{\rm disp}^2\sim (Gm_{\rm cl} \Omega)^{2/3} {t_{\rm coll}\over t_{\rm grav}}\sim  (Gm_{\rm cl} \Omega)^{2/3} \left({d_{\rm H}\over l_{\rm cl}}\right)^2.
\label{eq-vdisp2}
\end{equation}

The existence of this dispersion velocity has two consequences. First, it corresponds to a horizontal random walk, associated with an effective kinematic viscosity (Goldreich \& Tremaine 1978; Stewart \& Kaula 1980)
\begin{equation}
\nu_{\rm eff} \sim {V_{\rm disp}^2\over \Omega}{n_{\rm grav}\over n_{\rm grav}^2+1}
\label{eq-nueff}
\end{equation}
where $n_{\rm grav}=\left(d_{\rm cl}\over {R_{\rm H}}\right)^2$ is the number of gravitational interactions during one rotation period. It leads to an accretion rate of
\begin{equation}
\dot{M}= 3 \pi \Sigma_{\rm av} \nu_{\rm eff} \sim 3 \pi \Sigma_{\rm av} {V_{\rm disp}^2\over \Omega}{n_{\rm grav}\over n_{\rm grav}^2+1}
\label{eq-accretionrate}
\end{equation}
where $ \Sigma_{\rm av}$ is the average surface density.

Second, when it is an isotropic velocity dispersion (if the motions of the clouds are not restricted to a 2-dimensional structure, as is the case when the size of the clouds is smaller than the scale height of the disc), it intervenes in the hydrostatic equilibrium. One uses then the following prescription, which corresponds to replacing the thermal pressure in the hydrostatic equilibrium by the dynamical pressure:
 \begin{equation}
H\sim A {V_{\rm disp}\over \Omega}{1\over \sqrt{1+\zeta}},
\label{eq-scale height}
\end{equation}
where $A$ is a factor of order unity. The mass of the disc writes:
 \begin{equation}
M_{\rm disc}= \pi  R^2  \Sigma_{\rm av} \sim \pi R^2  {m_{\rm cl}^3\over d_{\rm cl}^2}\ {\rm max}\left[\left({H\over d_{\rm cl}}\right)^2,1\right].
\label{eq-massdisc1}
\end{equation}

We do not specify the heating mechanism, where we do not know what proportion is transformed into heat and radiated. As in the case of the continuous disc, we parametrize the heat flux by an effective temperature depending on the radius. Note that, if the coverage factor of the disc is smaller than unity, it should be taken into account as a factor multiplying the emitted flux in Eq. \ref{eq-lum-em}. 

To these ``macroscopic" equations, one should add the ``microscopic" ones corresponding to the state of the clouds. 
Equation \ref{eq:Fradbis} between the physical temperature and $T_{\rm eff}$ is still valid; but for the computation of the optical thickness, one must now take into account the number of clouds on a given line of sight, which depends on the coverage factor of the clouds $(l_{\rm cl}/ d_{\rm cl})^2 (H/d_{\rm cl})$. Equation \ref{eq:taubis} thus becomes
\begin{equation}
\tau=\rho H \kappa \  {\rm max}\left[\left({l_{\rm cl}\over d_{\rm cl}}\right)^2{H\over d_{\rm cl}},1\right].
\label{eq:tauter}
\end{equation}

\subsection{Mass transport by cloud collisions: applications}

\begin{figure}
\begin{center}
\includegraphics[width=9cm]{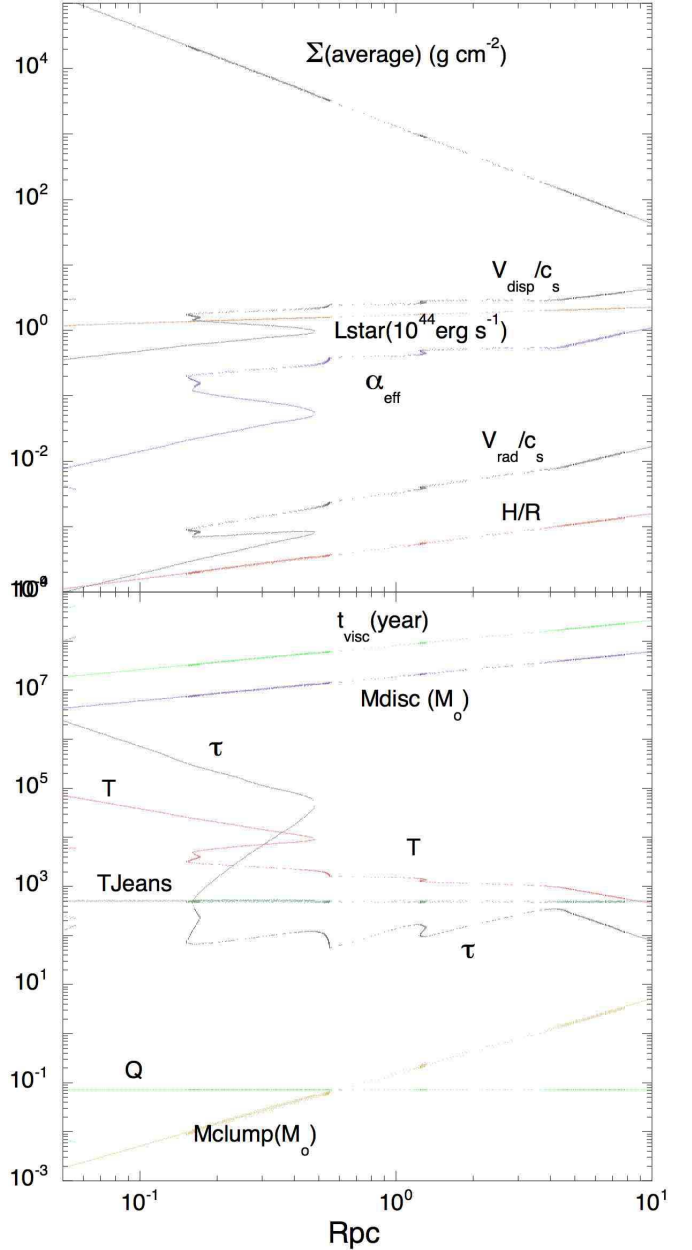}
\caption{Clumpy model for the luminous Seyfert spectrum shown in Fig. \ref{spectres}. The parameters of the model are  $M=10^8$  M$_{\odot}$,  Eddington ratio 0.1, $\alpha=0.01$, and a dependency of $T_{\rm eff}$ versus $R$:  $T_{\rm eff}\propto R^{-0.47}$, with $R_{\rm trans}$ equal to 0.5 10$^{-3}$ pc (Note that due a low value of $\alpha$ leads to a low value of $R_{\rm grav}$). The coverage factor is 0.2 and the ratio $l_{\rm cl}/d_{\rm cl}$ is equal to 0.2. The scarcity of points in some regions is due to the method used to sample the solutions.} 
\label{clumpydisc1}
\end{center}
\end{figure}

\begin{figure}
\begin{center}
\includegraphics[width=9cm]{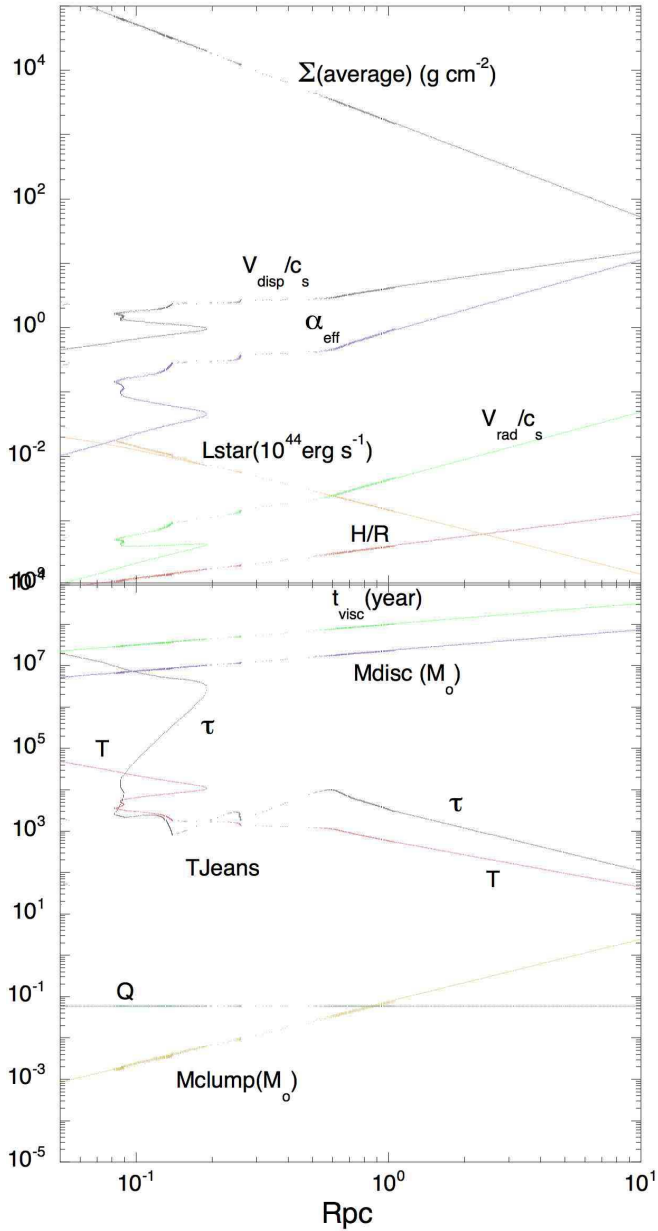}
\caption{Same as the previous figure, but now not trying to fit the luminous Seyfert spectrum. The parameters of the model are  $M=10^8$  M$_{\odot}$,  Eddington ratio 0.1, $\alpha=0.01$, and a dependency of $T_{\rm eff}$ versus $R$:  $T_{\rm eff}\propto R^{-0.75}$, with $R_{\rm trans}$ equal to 0.5 10$^{-3}$  pc. The coverage factor is 0.2 and the ratio $l_{\rm cl}/d_{\rm cl}$ is equal to 0.3. }     
\label{clumpydisc2}
\end{center}
\end{figure}

 \begin{figure*}
\begin{center}
\includegraphics[width=18cm]{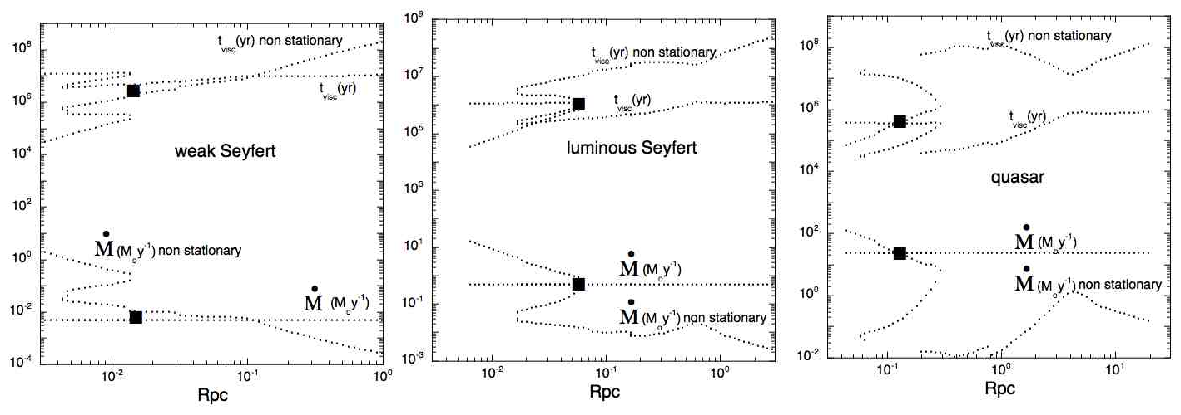}
\caption{ $t_{\rm visc}$ and  $\dot{M}$ as a function of the radius in the stationary (constant $\dot{M}$) and in the non stationary cases for the 3 previous models. The black dots mark the position of the transition between the $\alpha$-disc and the Q-disc in the stationary case.}     
\label{margnonstat}
\end{center}
\end{figure*}

 It is beyond the scope of this paper to develop a fully consistent model,  taking into account in particular the formation and the destruction of the clouds. Our aim is simply to show that a typical accretion rate can be achieved by the viscosity due to cloud interactions, for a steady disc made of small spherical clouds of same masses and average distances. 

\medskip
\noindent {\bf Case 1}

As for the continuous disc,  we would like to fit the typical spectrum of an AGN. In a first step we thus  compute the coverage factor. It intervenes simply as a multiplying factor in Eq. \ref{eq-lum-em}, so the value of $\gamma$ required to fit the observed spectrum stays the same as previously (0.47), and we also assume that the additional heating begins at  $R_{\rm trans}$. We adjust the viscosity coefficient $\alpha$ so that the $\alpha$-disc becomes gravitationally unstable at a relatively small radius to better fit the observations ($\alpha$=0.01). 
In the second step we impose the constant coverage factor found to fit the spectrum, as well as a constant ratio
 $l_{\rm cl}/d_{\rm cl}$. We solve the set of equations given in the appendix, and we compute the physical characteristics of the clouds (size, distance, temperature, density, etc) in order that the accretion rate be equal to the imposed value. 

To lighten the paper we give only the results concerning the luminous Seyfert. The  spectrum is shown in Fig. \ref{spectres}; it corresponds to a coverage factor equal to 0.2. The structure of the disc is displayed  in Fig. \ref{clumpydisc1} for a ratio  $l_{\rm cl}/d_{\rm cl}$ equal to 0.2. 

Here again, as for the continuous disc, there are multiple solutions between 0.1 and 0.4 pc, and consequently jumps in the radial structure due to the opacity gap. The transition between the $\alpha$-disc and the clumpy disc takes place at 0.03 pc, as shown by $\alpha_{\rm eff}=\alpha$ = 0.01. The disc is quite massive, reaching a mass close to that of the MBH at 10 pc, but it is very thin ($H/R$ between 10$^{-4}$ and  10$^{-3}$). As a consequence, $Q$ is lower than unity, but the clumps are gravitationally stable, since their masses are relatively low ($\le$ a few M$_{\odot}$) and the Jeans temperature is smaller than the clump temperature.
 The viscous time is long, typically a few 10$^7$ years, but stays compatible with the disc being stationary during the lifetime of the AGN.

Since the coverage factor is equal to 0.2, the disc absorbs only 20$\%$ of the external or internal heating flux, the rest being  radiated towards the observer. Owing to the additional heating required to fit the observed spectrum, it means about 10$\%$ of the bolometric luminosity should reach the clumpy disc. This value is high, but it is not unrealistic if the disc is warped.  This is indeed observed with the maser spots of NGC 4258, which actually constitutes a kind of ``clumpy disc".

\medskip
\noindent{\bf Case 2}

We can also consider another type of solution where the emission of the clumpy disc does not account for  the infrared band. 
 The observed infrared spectrum should then be provided by another source than the accretion disc, such as hot dust in the torus.  We assume a radius dependency 
$T_{\rm eff}\propto R^{-3/4}$, intermediate between a uniformly distributed source ($\gamma=1/2$), and a central point source at a small height above the disc ($\gamma=1$), and $\alpha$=0.01 for comparison with the previous model. 

The results for the luminous Seyfert are displayed in Fig. \ref{clumpydisc2}. The covering factor is 0.2, and the ratio  $l_{\rm cl}/d_{\rm cl}$ is equal to 0.3. The main difference from the previous model is the amount of heating, which now corresponds to only 0.1$\%$ to 1$\%$ of the bolometric luminosity.  The clumps are gravitationally unstable for a radius larger than 1 pc, so this clumpy disc cannot survive beyond this radius.   

\medskip
Though they can give the required accretion rate, clumpy discs are not satisfactory because they are quite sensitive to the coverage factor and to the ratio of the size to the distance of the clumps, for which there is neither physical justification nor observational constraints.  This is why we now discuss non-stationary discs. 

 \section{Non stationary discs in AGN}
  
 A solution allowing accretion without too much star formation in AGN and quasars consists in an intermittent process comparable to what is operating in the GC, i.e. successive waves of accretion lasting less than the viscous time of the outer disc.  There would thus be no need for another mechanism of transport than the viscous one. The accretion rate is given by Eq. \ref{eq:nuviscc}. 
  
 The problem with the initial state of the disc has already been mentioned about the GC: Nayakshin (2006) assumes that that it settles as a marginally stable disc. We make the same assumption. 
Then we have to define the non viscous heating.  There are  two options: either it is absent and appears only when the stars begin to form (cf. Nayakshin 2006), or it is present at the onset of each episode, for instance if it is given by a previous generation of stars. Since massive stars last for about 10$^7$ years, this is quite plausible. We adopt here the second hypothesis, which allows us to impose the resulting spectrum like in the study of stationary disc. As a result, the radial structure is  the same as in the case of a stationary disc, with only the accretion rate and the viscous time different. 

Figure \ref{margnonstat} compares $t_{\rm visc}$ and  $\dot{M}$ in the stationary and in the non stationary cases for the 3 previous models. The accretion rate in the non stationary case decreases outwards where it is much smaller than in the stationary case. But it is interesting to note that it is {\it equal to the stationary accretion rate} at the transition radius between the $\alpha$-disc and the Q-disc in the stationary case. At this position, the viscous time is about 10$^6$ years, while the non stationary viscous time is longer in the outer regions of the disc. Thus, if the duration of the accretion episodes is near 10$^6$ years, the inner $\alpha$-disc will be in viscous equilibrium and stationary during the accretion episode, while the outer disc will not be out in equilibrium. This is an interesting configuration, since {\it the emitted spectrum during this episode will be similar to the stationary one, i.e. close to the AGN spectrum}. 

That we do not know the initial state of the disc and that there is presently no physical basis for the marginal state hypothesis are not too serious. Actually, the disc can be in any other initial state: it must only stay optically thick to UV radiation, to be able to reprocess it into the infrared band and emit a spectrum in agreement with the observations. Even without any additional momentum transport,  it could be easily
gravitationally stable ($Q > 1$) owing to the stellar heating. In contrast, the flow can be gravitationally unstable when it starts up, and since stars are formed and begin to accrete after only 10$^{3-4}$ years, it becomes rapidly stabilized and able to reprocess the stellar radiation for the rest of the accretion episode. For each of these situations, we could give some numerical examples, but we avoid this here to keep this paper from being too long. 

  It is interesting to see the effect on non stationary discs of the thermal instability observed in a small region of stationary discs. If the accretion rate increases in the outer annulus of this region, and if the initial representative point is on the left stable curve  (cf. Fig. 5), the annulus will evolve through equilibrium states of increasing $\dot{M}$ into a viscous timescale. When $\dot{M}$ reaches the value corresponding to the maximum of this stable solution, the annulus will pass suddenly (into a thermal time) to the right stable equilibrium solution, triggering a strong perturbation (non Keplerian motions, shock waves, etc) which will  rapidly affect the whole unstable region. Inversely, if the accretion rate decreases, when it reaches the minimum values of the right stable curve of Fig. 5, the representative point will jump suddenly to the left stable curves, so one would expect that the unstable region answers any perturbation of the accretion rate more rapidly than the rest of the disc.  
 
Intermittent accretion has implications for the growth of quasars: it means that the ``duty cycle" represents only a small fraction of the cosmological time, as the ratio of the duration of the accretion episodes to the interval between two episodes is less than, say 10$\%$. For luminous (and therefore massive) high-redshift quasars, it is certainly too small. We also mentioned that it would contradict the continuity observed in radio-loud quasars between the large radio structure and the jets on the parsec scale. For local Seyfert nuclei, it is compatible with the small proportion of Seyfert galaxies among all galaxies (1$\%$). For a nucleus triggered by an interaction with another galaxy that lasts about 10$^8$ years, it means that it will only be active during 10$^7$ years. This implies that all galaxies should undergo at least 10 major interactions during their life to account for the observed number of Seyfert.

In summary, it seems that intermittent accretion could easily account for both stellar formation and accretion onto the black holes in Seyfert galaxies, while at the same time preserving an AGN spectrum comparable to the observations, but it does not solve the problem of high redshift and radio-loud quasars.
  
\section{Conclusion}

There are similarities between the Galactic Center and luminous AGN: in both cases, accretion should take place and build a massive disc at about one parsec from the MBH. This disc is gravitationally unstable and prone to stellar formation.  For the Galactic Center, the duration of the episode of accretion / star formation is too short for the disc to reach a stationary state. The suggestion of Nayakshin (2006) that the disc is maintained in a stable state by stellar heating can work if there is a mechanism forming a thin Keplerian disc in a time shorter than 10$^6$ years. If the process is identical to that acting in LLAGN, this implies that it should be repeated every 10$^7$ years and that there should be a permanent a population of massive hot stars very near the MBH around quiescent MBHs and in LLAGN. This is indeed observed in M31 and in the GC, and it begins to be observed within ten parsecs in some Seyfert galaxies (Davies et al. 2007).
	
Our purpose in this paper was to study the accretion rate / star formation process in quasars and AGN and to show that it can produce a spectrum similar to what is observed in AGN.
  We first considered the stationary case. To deal with the main problem for stationary discs, the transport of angular momentum, we focused on two mechanisms related specifically to the gravitational instability leading to the formation of clumps and to star formation, through supernovae explosions in a continuous disc or cloud collisions in a clumpy disc. 
 
We studied first the case of a continuous disc. Since the lifetime of the disc is longer than the formation and growth of the stars,  the rate of star formation should be maintained within modest limits so that the gaseous disc is not  immediately destroyed. It implies that  during a time close to or longer than the viscous time, the disc should be maintained in a state of marginal stability (a ``Q-disc"). 
However, it then requires additional heating and additional transport of angular momentum.
Starting from the observations, we studied the case of a disc accreting at a rate corresponding to the luminosity of the MBH. For the additional heating, what is provided by the newly formed massive stars can easily account for the spectral distribution of AGN; and at the same time it contributes to maintaining the disc in a marginally stable state. However we have shown that the transport provided by the supernovae issued from the massive star formation is not sufficient for producing the required extra amount of  momentum transport, except possibly in the weak Seyfert case. For quasars and luminous Seyfert, it is sufficient only at distances exceeding 10 pc. We have also shown that the Q-discs are perturbed by the existence of big jumps in their structure, due to the variations in the opacity with the temperature. It is probable that these jumps induce shockwaves and supersonic turbulence.

Then we examined a simple model of clumpy disc where the transport of angular momentum is provided by the interactions between the clouds. We showed that this transport is very efficient and can produce the required accretion rate, provided that small clumps are formed and maintained inside the disc. These clumps are hot enough to be gravitationally stable. The observed spectrum can be accounted for by the model, provided that the disc receives about 10$\%$ of the total luminosity of the object. The accretion rate can also be easily maintained if the disc is not irradiated  as much, but in this case its infrared emission is negligible. However, this model is unsatisfactory as the parameters (size and distance of the clumps) are tightly constrained without any physical justification.

Finally we considered intermittent accretion similar to that in the GC. If the duration of the accretion episodes is of the order of 10$^6$ years, the outer disc is not stationary and not in viscous equilibrium, in contrast to the inner disc, whose accretion rate corresponds to the luminosity of the objects. As a result the spectrum emitted by the inner and the outer discs agree with the observations. This model could work for Seyfert galaxies, but neither for luminous high-redshift quasars (because the duty cycle would be too short), nor for radio-loud quasars (because it does not ensure the required continuity between the large-scale and small-scale structures). 

A lot of work remains to be done. First, the structure of the ``jumps" in radial structure due to the opacity gap should be examined in detail, since they are present in all kinds of discs, whether gravitationally stable or unstable, stationary or not, clumpy or not, and it is  possible that the turbulence generated by these jumps has an impact on the angular momentum transport. Second, the clumpy disc deserves a thorough investigation, as it could be an interesting solution for the outer regions. And finally it would be necessary to study a genuine time-dependent model taking into account the evolution of star formation and its feed-back on the disc, as it could certainly represent a solution for Seyfert nuclei.

\begin{acknowledgements}

\end{acknowledgements}

\appendix

\section{Equations of the continuous disc structure}

We recall here the set of equations for solving the radial structure of the $\alpha$-disc. Some of them have already been given in the text, but we prefer to rewrite them to get a fully consistent ensemble. We do not redefine the parameters already mentioned in the text.

The equation of hydrostatic equilibrium writes as
\begin{equation}
c_{\rm s}^2 = \Omega^2 H (1+\zeta)
\label{eq:hydrostatic}
\end{equation}
with
\begin{equation}
\zeta=4\pi G \rho/\Omega^2,
\label{eq:zeta}
\end{equation}
\begin{equation}
c_{\rm s} = {P_{\rm gas}+ P_{\rm rad} \over \rho}, \ \ {\rm and}
\label{eq:sound}
\end{equation}
\begin{equation}
P_{\rm gas} = \frac{\rho k T }{\mu m_{\rm H}},
\label{eq:Pgas}
\end{equation}
where  $\mu$ is the mean mass per 
particle, $m_{\rm H}$ the proton mass. Then
\begin{equation}
 P_{\rm rad}={\tau\sigma T_{\rm eff}^4 \over 2c},
\label{eq:Prad}
\end{equation}
where $\tau$ is the Rosseland mean opacity:
\begin{equation}
\tau=\rho H \kappa ={\Sigma \kappa \over 2},
\label{eq:tau}
\end{equation}
where $\kappa$ is the opacity coefficient in g$^{-1}$ cm$^2$. 

The diffusion equation writes as
\begin{equation}
\sigma T_{\rm eff}^4={\sigma T^4\over 3\tau/8+1/2+1/4\tau}.
\label{eq:Frad}
\end{equation}

The energy balance between the local viscous dissipation and the
radiative cooling requires
\begin{equation}
F_{\rm rad} =\sigma T_{\rm eff}^4= \frac{3 \Omega^2 \dot{M}f(R)}{8 \pi}.
\label{eq:eb}
\end{equation}

When there is an additional non viscous heating, this equation is simply replaced by $T_{\rm eff}\propto R^{-\gamma}$, ensuring continuity of $T_{\rm eff}$ at the transition radius. 

The $\alpha$-prescription defines the
viscosity coefficient as
\begin{equation}
\nu = \alpha \sqrt{\frac{P}{\rho}} H,
\label{eq:alpha}
\end{equation}
$\nu$ being the kinematic viscosity.
And finally the equation accounting for the transport of angular momentum is
\begin{equation}
\frac{\dot{M} f(R)}{3 \, \pi} = 2 \nu \rho H.
\label{eq:nuvisc}
\end{equation}

When solved self-consistently with the appropriate 
functions for $\kappa(\rho,T)$ and 
$\mu(\rho,T)$, these equations yield the $\alpha$-disk structure. 

In the Q-disc, one has
\begin{equation}
Q = {\Omega c_{\rm s}\over \pi G\Sigma}\ ={\Omega^2 \over 2 \pi G \rho} \sqrt{1+\zeta}=1,
\label{eq:Q-disc}
\end{equation}
which requires suppressing one of the equations in the previous set. As explained in the text, Sirko \& Goodman (2003) suppress the energy-balance equation \ref{eq:eb}, while we suppress the equation for the moment transport \ref{eq:nuvisc}. However, as explained in Section 3.2,  we take the viscous transport into account by allowing a fraction of $\dot{M}$ - which also contributes to the heating - be due to the viscous transport. The set of equations is solved by two different iterations for the $\alpha$- and for the Q- discs. 

\begin{figure}
\begin{center}
\includegraphics[width=9cm]{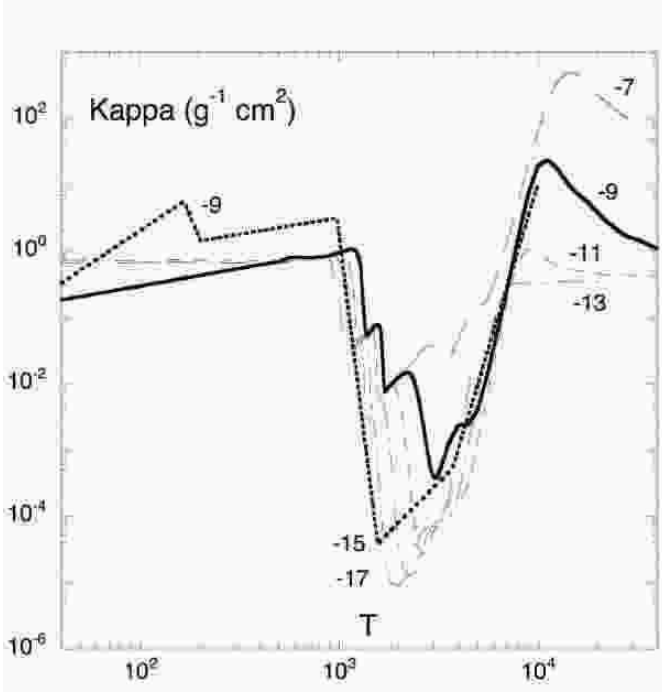}
\caption{Ferguson et al. opacities used in this paper as a function of the temperature. The curves are labeled with the log of the density in g cm$^{-3}$. For comparison the two thick  lines show the Lin \& Bell (dotted line) and the Ferguson et al. (solid line) opacities for the same density $\rho$=10$^{-9}$  g cm$^{-3}$.}     
\label{opacite}
\end{center}
\end{figure}

Since we are dealing with a gas between a few hundred and a few thousand degrees in the  Q-disc of interest for us, we take $\mu=1$, a value intermediate between 2.4, that is valid for a fully molecular gas, and 0.6 valid for a fully-ionized gas. For the opacities, we use the tables of Ferguson et al. (2005) for solar heavy-element abundances (set g7.02.tron, with the Grevesse \& Noels abundances 1993, X=0.7 and Z= 0.02). These opacities are displayed in Fig. \ref{opacite} and compared to those of Bell \& Lin (1994) generally used in these problems, which should be extrapolated for low densities.

\section{Angular momentum provided by supernovae}

Rozyzka et al.  assume
that the disc receives all the linear momentum projected on the mid plane of the disc (i.e. about $P_{\rm tot}$/2).   Thus the velocity reached by the remnant when it has expanded to a distance $\Delta R$ is
\begin{equation}
v_{\rm e}={P_{\rm tot}\over 2} {1\over \pi \Sigma \Delta R^2 }.
\label{eq-Roz1}
\end{equation} 
 
 The remnant is stopped when this velocity equals the shear velocity at the edge relative to the explosion, $\Omega \Delta R/2$. Combining this value with Eq. \ref{eq-Roz1} gives $\Delta R_{\rm final}$ and $v_{\rm final}$, and finally the angular momentum given to the disc (see Rozyzka et al. for more details):
\begin{equation}
\Delta J =  {2\over 3\pi} P_{\rm tot} R.
\label{eq-Roz2}
\end{equation} 
From this value, Rozyzka et al. deduced that a rate of supernova explosions of 10$^{-4}$ to 10$^{-3}$ per year corresponds to the viscous transport in an $\alpha$-disc having $\alpha$=0.1 to 1.  

There is a big 
difference from 
CZ99. The Rozyczka et al. computation  does not depend on the scale height, 
but only on the surface
density which is given a priori. Taking the 3-dimensional aspect into account, CZ99 assume
that the disk only receives the momentum corresponding to  the velocity 
vectors inclined by less than
$H/R$ to the mid-plane. Calling $\theta_{\rm max}$ the angle between the equatorial plane and the point where the remnant breaks through the disc surface, one gets the angular momentum supplied by a supernova,
\begin{equation}
\Delta J = {P_{\rm tot}\over 4 \pi} \int_{-\theta_{\rm max}}^{+\theta_{\rm max}} cos(\theta) 2\pi cos(\theta) d\theta , 
\label{eq-angular-momentum-3}
\end{equation} 
or
\begin{equation}
\Delta J = {P_{\rm tot}\over 2} [\theta_{\rm max} + {sin(2\theta_{\rm max})\over 2}]. 
\label{eq-angular-momentum-4}
\end{equation} 
For small angles, this expression can be approximated by
\begin{equation}
\Delta J \sim {3\over 2\pi} P_{\rm tot} R {H\over R_{\rm  s,max}}, 
\label{eq-angular-momentum-bis}
\end{equation} 
where  $R_{\rm  s,max}$ is the maximum extension of the supernova before it breaks vertically out of the disc: 
\begin{equation}
R_{\rm  s,max} \sim \left({3P_{\rm tot}\over 4\pi\rho\Omega}\right)^{1/4}.
\label{eq-Rsmax1SNbis}
\end{equation} 
The angular momentum given by  Eq. \ref{eq-angular-momentum-bis} differs from that of Rozyczka et al. by a factor $9/4(H/R_{\rm  s,max})$,  which is generally much smaller than unity; so our estimation is more conservative than that of Rozyczka et al. and it requires a larger number of supernovae to achieve the same momentum transport. 

The rate of supernovae per decade of radius is finally given  by the relation:
\begin{equation}
 {\cal N}_{\rm SN}  \sim {2 \dot{M}_{\rm SN} R^2 \Omega \over \Delta J 
log_{10}e}.
\label{eq-NSN}
\end{equation} 

\section{Equations of the clumpy disc structure}

We give here the complete set of equations used for the radial structure of the  clumpy disc, though the majority of them have already been given in the text: 

\begin{equation}
V_{\rm disp}= (Gm_{\rm cl} \Omega)^{1/3}{d_{\rm H}\over l_{\rm cl}} .
\label{eq-vdispapp}
\end{equation}
\begin{equation}
m_{\rm cl}= l_{\rm cl} \rho^3 ,
\label{eq-mcloudapp}
\end{equation}
where $\rho$ is the density inside the clumps.
\begin{equation}
d_{\rm H}=R\left({m_{\rm cl}\over M_{\rm BH}}\right)^{1/3}.
\label{eq-dtapp}
\end{equation}
 \begin{equation}
M_{\rm disc}= \pi  R^2  \Sigma_{\rm av} .
\label{eq-massdisc1app}
\end{equation}
 \begin{equation}
\Sigma_{\rm av}= {m_{\rm cl}^3\over d_{\rm cl}^2}\ {\rm max}\left[\left({H\over d_{\rm cl}}\right)^2,1\right] .
\label{eq-gsigmamoyapp}
\end{equation}
 \begin{equation}
H\sim A {V_{\rm disp}\over \Omega}{1\over \sqrt{1+\zeta_{\rm av}}},
\label{eq-scale heightapp}
\end{equation}
where $A$ is a factor of order unity, and $\zeta_{\rm av}$ is the parameter $\zeta$ averaged on the disc:
 \begin{equation}
\zeta_{\rm av}={4\pi G\rho_{\rm av}\over \Omega^2},
\label{eq-zetamoyapp}
\end{equation}
with 
 \begin{equation}
\rho_{\rm av}={\Sigma_{\rm av}\over 2H}.
\label{eq-rhomoyapp}
\end{equation}
\begin{equation}
\dot{M}= 3 \pi \Sigma_{\rm av} {V_{\rm disp}^2\over \Omega}{\left(d_{\rm cl}\over {R_{\rm H}}\right)^2\over \left(d_{\rm cl}\over {R_{\rm H}}\right)^4+1}.
\label{eq-accretionrateapp}
\end{equation}
\begin{equation}
\tau=l_{\rm cl}\kappa (\rho,T)\rho \ {\rm max}\left(1,{l_{\rm cl}^2H\over d_{\rm cl}^3}\right).
\label{eq-tauapp}
\end{equation}
\begin{equation}
\sigma T_{\rm eff}^4={\sigma T^4\over 3\tau/8+1/2+1/4\tau}
\label{eq:Fradapp}
\end{equation}
where $T_{\rm eff}=R^{-\gamma}$, with $\gamma$ imposed.
These equations are completed by the two  conditions:
\begin{equation}
{l_{\rm cl}\over H}= {\rm given\ constant 1}
\label{eq-cond1app}
\end{equation}
\begin{equation}
{d_{\rm cl}\over H}= {\rm given\ constant 2}
\label{eq-cond2app}
\end{equation}
or equivalently given coverage factor, and given $l_{\rm cl}/d_{\rm cl}$.
We thus have a set of 13 equations with 13 unknowns, which can be solved by an iteration procedure. We also compute the Jeans temperature of the clumps, $T_{\rm Jeans}$, to check their gravitational stability:
\begin{equation}
 T_{\rm Jeans}=\left(10^{23}\mu m_{\rm cl}\sqrt{\rho}\right)^{2/3}.
\label{eq:Tjeansapp}
\end{equation}


\begin{thebibliography}{}

\bibitem{} Artymowicz, P., Lin, D.N., \& Wampler, E.J. 1993, ApJ 409, 592 
\bibitem[Barthel(1989)]{1989ApJ...336..606B} Barthel, P.~D.\ 1989, \apj, 
336, 606 
\bibitem{} Begelman, M.C., Frank J., Shlosman I. 1989, in ``Theory 
of Accretion disks", eds. F. Meyer et al., Kluwer Academic Publishers
\bibitem[Bell \& Lin(1994)]{1994ApJ...427..987B} Bell, K.~R., \& Lin, 
D.~N.~C.\ 1994, \apj, 427, 987 
\bibitem[Bender et al.(2005)]{2005ApJ...631..280B} Bender, R., et al.\ 
2005, \apj, 631, 280 
\bibitem[Bower et al.(2005)]{2005ApJ...618L..29B} Bower, G.~C., Falcke, H., 
Wright, M.~C., \& Backer, D.~C.\ 2005, \apjl, 618, L29 
\bibitem[Clavel et al.(1989)]{1989ApJ...337..236C} Clavel, J., Wamsteker, 
W., \& Glass, I.~S.\ 1989, \apj, 337, 236 
\bibitem[Christopher et al.(2005)]{2005ApJ...622..346C} Christopher, M.~H., 
Scoville, N.~Z., Stolovy, S.~R., \& Yun, M.~S.\ 2005, \apj, 622, 346 
\bibitem[Collin \& Hur{\'e}(1999)]{1999A&A...341..385C} Collin, S., \& 
Hur{\'e}, J.-M.\ 1999, \aap, 341, 385 
\bibitem[Collin \& Zahn(1999)]{1999Ap&SS.265..501C} Collin, S., \& Zahn, 
J.-P.\ 1999a, \apss, 265, 501 
\bibitem[Collin \& Zahn(1999)]{1999A&A...344..433C} Collin, S., \& Zahn, 
J.-P.\ 1999b, \aap, 344, 433 (CZ99)
\bibitem[Cuadra et al.(2003)]{2003A&A...411..405C} Cuadra, J., Nayakshin, 
S., \& Sunyaev, R.\ 2003, \aap, 411, 405 
\bibitem[Cuadra et al.(2005)]{2005MNRAS.360L..55C} Cuadra, J., Nayakshin, 
S., Springel, V., \& Di Matteo, T.\ 2005, \mnras, 360, L55 
\bibitem[Cuadra et al.(2006)]{2006MNRAS.366..358C} Cuadra, J., Nayakshin, 
S., Springel, V., \& di Matteo, T.\ 2006, \mnras, 366, 358 
\bibitem{}Davies, R., Genzel, R., Tacconi, L., Mueller Sanchez, F., Sternberg, A. 2007,  ArXiv 
Astrophysics e-prints, arXiv:astro-ph/0612009
\bibitem[Deegan \& Nayakshin(2007)]{2007MNRAS.377..897D} Deegan, P., \& 
Nayakshin, S.\ 2007, \mnras, 377, 897 
\bibitem[Duschl et al.(2000)]{2000A&A...357.1123D} Duschl, W.~J., 
Strittmatter, P.~A., \& Biermann, P.~L.\ 2000, \aap, 357, 1123 
\bibitem[Duschl \& Britsch(2006)]{2006ApJ...653L..89D} Duschl, W.~J., \& 
Britsch, M.\ 2006, \apjl, 653, L89 
\bibitem[Eckart et al.(1999)]{1999A&A...352L..22E} Eckart, A., Ott, T., \& 
Genzel, R.\ 1999, \aap, 352, L22 
\bibitem[Ferguson et al.(2005)]{2005ApJ...623..585F} Ferguson, J.~W., 
Alexander, D.~R., Allard, F., Barman, T., Bodnarik, J.~G., Hauschildt, 
P.~H., Heffner-Wong, A., \& Tamanai, A.\ 2005, \apj, 623, 585 
\bibitem[Gammie et al.(1991)]{1991ApJ...378..565G} Gammie, C.~F., Ostriker, 
J.~P., \& Jog, C.~J.\ 1991, \apj, 378, 565 
\bibitem[Gammie(2001)]{2001ApJ...553..174G} Gammie, C.~F.\ 2001, \apj, 553, 
174 
\bibitem[Genzel et al.(2003)]{2003ApJ...594..812G} Genzel, R., et al.\ 
2003, \apj, 594, 812 
\bibitem[Gerhard(2001)]{2001ApJ...546L..39G} Gerhard, O.\ 2001, \apjl, 546, 
L39 
\bibitem{} Goldreich, P., Lynden-Bell, D. 1965 MNRAS 130, 97
\bibitem[Goldreich \& Tremaine(1978)]{1978Icar...34..227G} Goldreich, P., 
\& Tremaine, S.~D.\ 1978, Icarus, 34, 227 
\bibitem[Goodman(2003)]{2003MNRAS.339..937G} Goodman, J.\ 2003, \mnras, 
339, 937 
\bibitem[Greenhill et al.(1995)]{1995A&A...304...21G} Greenhill, L.~J., 
Henkel, C., Becker, R., Wilson, T.~L., \& Wouterloot, J.~G.~A.\ 1995, \aap, 
304, 21 
 \bibitem[Grevesse \& Noels(1993)]{1993oee..conf...14G} Grevesse, N., \& 
Noels, A.\ 1993, Origin and evolution of the elements: proceedings of a 
symposium in honour of H.~Reeves, held in Paris, June 22-25, 1992.~Edited 
by N.~Prantzos, E.~Vangioni-Flam and M.~Casse.~Published by Cambridge 
University Press, Cambridge, England, 1993, p.14
\bibitem[Haardt \& Maraschi(1991)]{1991ApJ...380L..51H} Haardt, F., \& 
Maraschi, L.\ 1991, \apjl, 380, L51 
\bibitem[Haardt \& Maraschi(1993)]{1993ApJ...413..507H} Haardt, F., \& 
Maraschi, L.\ 1993, \apj, 413, 507 
\bibitem[Ho et al.(1997)]{1997ApJ...487..568H} Ho, L.~C., Filippenko, 
A.~V., \& Sargent, W.~L.~W.\ 1997, \apj, 487, 568 
\bibitem[Hure et al.(1994)]{1994A&A...290...34H} Hur\'e, J.-M., 
Collin-Souffrin, S., Le Bourlot, J., \& Pineau des Forets, G.\ 1994, \aap, 
290, 34 
\bibitem[Hure(1998)]{1998A&A...337..625H} Hur\'e, J.-M.\ 1998, \aap, 337, 625
\bibitem[Jackson et al.(1993)]{1993ApJ...402..173J} Jackson, J.~M., Geis, 
N., Genzel, R., Harris, A.~I., Madden, S., Poglitsch, A., Stacey, G.~J., \& 
Townes, C.~H.\ 1993, \apj, 402, 173  
\bibitem[Jog \& Ostriker(1988)]{1988ApJ...328..404J} Jog, C.~J., \& 
Ostriker, J.~P.\ 1988, \apj, 328, 404 
\bibitem[Jog \& Ostriker(1989)]{1989ApJ...337.1035J} Jog, C.~J., \& 
Ostriker, J.~P.\ 1989, \apj, 337, 1035 
\bibitem[Kollatschny(2003)]{2003A&A...407..461K} Kollatschny, W.\ 2003, 
\aap, 407, 461 
\bibitem{} Kolykhalov, P.I., Sunyaev, R.A. 1980, Soviet Astron. 
Lett. 6, 357
\bibitem[Koratkar \& Blaes(1999)]{1999PASP..111....1K} Koratkar, A., \& 
Blaes, O.\ 1999, \pasp, 111, 1 
\bibitem[Kumar(1999)]{1999ApJ...519..599K} Kumar, P.\ 1999, \apj, 519, 599 
\bibitem[Levin \& Beloborodov(2003)]{2003ApJ...590L..33L} Levin, Y., \& 
Beloborodov, A.~M.\ 2003, \apjl, 590, L33 
\bibitem[Levin(2007)]{2007MNRAS.374..515L} Levin, Y.\ 2007, \mnras, 374, 
515 
\bibitem[Lin \& Pringle(1987)]{1987MNRAS.225..607L} Lin, D.~N.~C., \& 
Pringle, J.~E.\ 1987, \mnras, 225, 607 
\bibitem[Lodato \& Bertin(2001)]{2001A&A...375..455L} Lodato, G., \& 
Bertin, G.\ 2001, \aap, 375, 455 
\bibitem[Milosavljevi{\'c} \& Loeb(2004)]{2004ApJ...604L..45M} 
Milosavljevi{\'c}, M., \& Loeb, A.\ 2004, \apjl, 604, L45
\bibitem[Morris \& Nayakshin(2007)]{2007astro.ph..1047M} Morris, M., \& 
Nayakshin, S.\ 2007, ArXiv Astrophysics e-prints, arXiv:astro-ph/0701047 
\bibitem[Murray \& Chiang(1997)]{1997ApJ...474...91M} Murray, N., \& 
Chiang, J.\ 1997, \apj, 474, 91 
\bibitem[Nayakshin(2004)]{2004MNRAS.352.1028N} Nayakshin, S.\ 2004, \mnras, 
352, 1028 
\bibitem[Nayakshin \& Cuadra(2005)]{2005A&A...437..437N} Nayakshin, S., \& 
Cuadra, J.\ 2005, \aap, 437, 437 
\bibitem[Nayakshin \& Sunyaev(2005)]{2005MNRAS.364L..23N} Nayakshin, S., \& 
Sunyaev, R.\ 2005, \mnras, 364, L23 
\bibitem[Nayakshin(2006)]{2006MNRAS.372..143N} Nayakshin, S.\ 2006, \mnras, 
372, 143 
\bibitem[Nayakshin et al.(2006)]{2006MNRAS.366.1410N} Nayakshin, S., 
Dehnen, W., Cuadra, J., \& Genzel, R.\ 2006, \mnras, 366, 1410 
\bibitem[Nayakshin et al.(2007)]{2007MNRAS.tmp..517N} Nayakshin, S., 
Cuadra, J., \& Springel, V.\ 2007, \mnras, 517 
\bibitem[Nayakshin(2007)]{2007astro.ph..1150N} Nayakshin, S.\ 2007, ArXiv 
Astrophysics e-prints, arXiv:astro-ph/0701150 
\bibitem[Ozernoy et al.(1998)]{1998A&A...337..105O} Ozernoy, L.~M., 
Fridman, A.~M., \& Biermann, P.~L.\ 1998, \aap, 337, 105  
\bibitem{} Paczynski, B. 1978, AcA, 28, 91
\bibitem[Paumard et al.(2006)]{2006ApJ...643.1011P} Paumard, T., et al.\ 
2006, \apj, 643, 1011 
\bibitem[Peterson \& Wandel(1999)]{1999ApJ...521L..95P} Peterson, B.~M., \& 
Wandel, A.\ 1999, \apjl, 521, L95 
\bibitem[Quataert et al.(1999)]{1999ApJ...517L.101Q} Quataert, E., Narayan, 
R., \& Reid, M.~J.\ 1999, \apjl, 517, L101 
\bibitem[Quataert \& Gruzinov(2000)]{2000ApJ...539..809Q} Quataert, E., \& 
Gruzinov, A.\ 2000a, \apj, 539, 809 
\bibitem[Quataert \& Gruzinov(2000)]{2000ApJ...545..842Q} Quataert, E., \& 
Gruzinov, A.\ 2000b, \apj, 545, 842 
\bibitem[Revnivtsev et al.(2004)]{2004A&A...425L..49R} Revnivtsev, M.~G., 
et al.\ 2004, \aap, 425, L49 
\bibitem{} Rozyczka, M., Bodenheimer P.,  Lin D.ÊN.ÊC. 1995, MNRAS 
276, 597
\bibitem{} Shakura, N.I., Sunyaev, R.A. 1973, A\&A 24, 337
\bibitem{} Shlosman, I., Begelman, M. C. 1989, ApJ 341, 685 
\bibitem{} Shlosman, I., Frank, J., Begelman, M.C. 1989, Nature 338, 45
\bibitem[Sirko \& Goodman(2003)]{2003MNRAS.341..501S} Sirko, E., \& 
Goodman, J.\ 2003, \mnras, 341, 501 
\bibitem[Stewart \& Kaula(1980)]{1980Icar...44..154S} Stewart, G.~R., \& 
Kaula, W.~M.\ 1980, Icarus, 44, 154 
\bibitem[Thompson et al.(2005)]{2005ApJ...630..167T} Thompson, T.~A., 
Quataert, E., \& Murray, N.\ 2005, \apj, 630, 167 
\bibitem{} Toomre, A. 1964, ApJ 139, 1217
\bibitem[Vollmer \& Duschl(2001)]{2001A&A...367...72V} Vollmer, B., \& 
Duschl, W.~J.\ 2001, \aap, 367, 72 
\bibitem[Vollmer \& Beckert(2002)]{2002A&A...382..872V} Vollmer, B., \& 
Beckert, T.\ 2002, \aap, 382, 872 
\bibitem{} Wang, B., \& Silk, J. 1994, ApJ 427, 759


\end{thebibliography}
\end{document}